\documentclass[letterpaper,twocolumn,prl,
aps,showpacs,superscriptaddress,floatfix]{revtex4-2}

\usepackage[latin1]{inputenc}
\usepackage{bbm}
\usepackage{bm}
\usepackage[usenames]{color}
\usepackage{multirow}
\usepackage{amssymb}
\usepackage{amsbsy}
\usepackage{mathtools}
\usepackage{amsmath}
\usepackage{stmaryrd}
\usepackage{graphicx}
\usepackage{epsfig}
\usepackage{placeins}
\usepackage{bbold}
\usepackage{braket}
\usepackage{blindtext}
\usepackage[colorlinks,linkcolor=blue,citecolor=blue,urlcolor=blue]{hyperref}

\usepackage{scalerel}
\usepackage{tikz}
\usetikzlibrary{calc}
\usetikzlibrary{patterns}
\usetikzlibrary{svg.path}
\definecolor{orcidlogocol}{HTML}{A6CE39}
\tikzset{
	orcidlogo/.pic={
		\fill[orcidlogocol] svg{M256,128c0,70.7-57.3,128-128,128C57.3,256,0,198.7,0,128C0,57.3,57.3,0,128,0C198.7,0,256,57.3,256,128z};
		\fill[white] svg{M86.3,186.2H70.9V79.1h15.4v48.4V186.2z}
		svg{M108.9,79.1h41.6c39.6,0,57,28.3,57,53.6c0,27.5-21.5,53.6-56.8,53.6h-41.8V79.1z M124.3,172.4h24.5c34.9,0,42.9-26.5,42.9-39.7c0-21.5-13.7-39.7-43.7-39.7h-23.7V172.4z}
		svg{M88.7,56.8c0,5.5-4.5,10.1-10.1,10.1c-5.6,0-10.1-4.6-10.1-10.1c0-5.6,4.5-10.1,10.1-10.1C84.2,46.7,88.7,51.3,88.7,56.8z};
	}
}

\newcommand\orcid[1]{\!%
	\href{https://orcid.org/#1}{%
		\mbox{%
			\scaleto{%
				\begin{tikzpicture}[yscale=-1,transform shape]
				\pic{orcidlogo};
				\end{tikzpicture}
			}{8pt}%
		}%
	}%
}

\makeatletter

\begin{document}
	\title{Emergence of unitary symmetry  of microcanonically truncated
		operators in chaotic quantum systems}
	
	\author{Jiaozi Wang~\orcid{0000-0001-6308-1950}}
	\affiliation{Department of Mathematics/Computer Science/Physics, University of Osnabr\"uck, D-49076 
		Osnabr\"uck, Germany}
	
	\author{Jonas Richter~\orcid{0000-0003-2184-5275}}
	\affiliation{Department of Physics, Stanford University, Stanford, CA 94305, 
		USA}
	\affiliation{Institut f\"ur Theoretische Physik, Leibniz 
		Universit\"at Hannover, 30167 Hannover, Germany}

	\author{Mats H. Lamann}
	\affiliation{Department of Mathematics/Computer Science/Physics, University of Osnabr\"uck, D-49076 
		Osnabr\"uck, Germany}
	
	\author{Robin Steinigeweg~\orcid{0000-0003-0608-0884}}
	\affiliation{Department of Mathematics/Computer Science/Physics, University of Osnabr\"uck, D-49076 
		Osnabr\"uck, Germany}
	
	\author{Jochen Gemmer}
	\affiliation{Department of Mathematics/Computer Science/Physics, University of Osnabr\"uck, D-49076 
		Osnabr\"uck, Germany}

	\author{Anatoly Dymarsky~\orcid{0000-0001-5762-6774}}
	\affiliation{Department of Physics, University of Kentucky, 
		Lexington, Kentucky, USA, 40506}

	\date{\today}
	
	\begin{abstract}
		We study statistical properties of matrix elements of observables written in the energy eigenbasis  
		and truncated to small microcanonical  windows. 
		We present numerical evidence indicating that for all few body operators  in chaotic many-body
		systems,  truncated below certain energy scale, 
		collective statistical properties of matrix elements exhibit emergent unitary symmetry.  
		Namely, we show that below certain scale the spectra of the truncated operators exhibit  universal 
		behavior,  matching  our analytic predictions, which are  numerically testable for system sizes beyond exact  diagonalization.
		We discuss operator and system-size dependence of the energy scale of emergent unitary symmetry
		and put our findings in context of previous works exploring emergence of random-matrix 
		behavior at small energy scales.
		
	\end{abstract}
	
	\maketitle


	{\it Introduction. }
	The eigenstate thermalization 
	hypothesis (ETH) \cite{Deutsch,Srednicki94} provides a microscopic explanation for the emergence of 
	thermodynamic behavior in isolated quantum systems. In the spirit of random-matrix theory, the ETH asserts 
	individual energy eigenstates with similar energies are physically equivalent. Accordingly, matrix elements of observables ${ O}_{mn}$ written in the energy eigenbasis can be described statistically. Qualitatively this leads to a picture where energy eigenstates confined to a sufficiently narrow microcanonical energy window can be randomly reshuffled without changing the statistics of ${ O}_{mn}$. This picture of ETH based on typicality was advocated in  \cite{Review,PhysRevE.97.012140,PhysRevLett.129.170603}. 
	
{As a stronger statement, parallel to the
Bohigas-Giannoni-Schmit conjecture for the energy spectrum \cite{Bohigas}, for
quantum chaotic systems one may expect that matrix elements ${ O}_{mn}$ at
sufficiently small scales exhibit full Random-Matrix-Theory universality. A
more careful analysis shows that an {\it uncorrelated}, i.e.\ Gaussian RMT
description, can only apply at very small scales \cite{PhysRevE.102.042127,
PhysRevLett.128.180601, PhysRevLett.128.190601}, as the correlations between
matrix elements are inevitable and important to describe various aspects of
quantum dynamics \cite{PhysRevLett.128.190601, FoiniPRE,Chan:2018fsp,
Murthy2019, Brenes2021}. An important step is the evidence that ${O}_{mn}$ can
be described by a rotationally invariant ensemble, i.e.\ general orthogonal or
unitary-invariant RMT, that assumes cross-correlations of ${O}_{mn}$
\cite{PhysRevLett.123.260601}. In this Letter, we clarify and further establish
this picture by first analytically deriving an infinite set of signatures of
unitary/orthogonal invariance, and then by testing them numerically. Crucially,
we identify relevant energy scale(s) marking the onset of full or partial
unitary symmetry. Our numerical approach is based on quantum typicality, and
hence applicable to system sizes substantially beyond the reach of exact
diagonalization. We confirm that the behavior is consistent with unitary
symmetry for all operators in chaotic quantum systems, while for integrable
models these signatures are notably absent \cite{SuppMat}.}

	{\it Unitary symmetry of microcanonically-truncated operators.} 
	{We consider an observable $O$ satisfying ETH, with its smooth diagonal part subtracted ${\cal O}_{mn} \equiv O_{mn}-O(E)\delta_{mn}$. It is then projected (microcanonically truncated) onto a narrow energy window of width $\Delta E$,
	\begin{equation}
	\label{OdE}
	{\cal O}_{\Delta E} = P_{\Delta E}\,  {\cal O}\, P_{\Delta E}\ ,  
	\end{equation}
	where we introduce the projector,
	\begin{equation}\label{Eq::EFilter}
	P_{\Delta E} = \sum_{|E_m-E_0| < \Delta E/2} \ket{E_m}\bra{E_m}\ , 
	\end{equation}
	and $d={\rm Tr}(P_{\Delta E})$ will denote the number of states within the microcanonical window. }
	We expect that for a sufficiently small scale  $\Delta E \le \Delta E_U$, operator ${\cal O}_{\Delta E}$ will exhibits emergent unitary symmetry $U(d)$\cite{Symmetry}, which will impose  constraints on correlations of the matrix elements as well as on the spectrum of ${\cal O}_{\Delta E}$ as a function of $\Delta E$.
	The idea of studying spectral properties of ${\cal O}_{\Delta E}$ in conjunction with their $\Delta E$ dependence was put forward in \cite{PhysRevE.99.010102}, with more detailed studies to follow \cite{PhysRevB.99.224302,PhysRevLett.128.190601,PhysRevE.102.042127,PhysRevLett.128.180601, pappalardi2023microcanonical, iniguez2023microcanonical}. A convenient way
	to probe the spectrum of the microcanonically truncated operator is through its moments 
	\begin{equation}\label{Eq::Mom}
	{\cal M}_k(\Delta E) = \frac{\text{Tr}[({\cal O}_{\Delta E})^k]}{d(\Delta E)}\ , 
	\end{equation}
	which can be combined into free cumulants $\Delta_k= \Delta_k(\Delta E)$ \cite{PhysRevLett.129.170603}, defined through iterative relation
	\begin{equation}\label{Eq::cumulants}
	\Delta_{k} = {\cal M}_{k}-\sum_{j=1}^{k-1}\Delta_{j}\sum_{a_{1}+a_{2}+\cdots a_{j}=k-j}{\cal M}_{a_{1}}\cdots{\cal M}_{a_{j}}\ .
	\end{equation}
	We show in the Supplemental Material \cite{SuppMat} that  unitary symmetry 
	requires $\Delta_{k}(\Delta E) \propto (d(\Delta E))^{k-1}$. When $\Delta E_U$ is much smaller than the effective temperature, which is always the case for spatially-extended systems,
	density of states within the microcanonical window is approximately constant,  $d(\Delta E)/d(\Delta E_U)=\Delta E/\Delta E_U$, leading to 
	{\begin{equation}\label{eq-Deltak-DE}
		\Delta_{k}(\Delta E)=\lambda_{k}\,\Delta E^{k-1} 
		\end{equation}
for $\Delta E \leq \Delta E_U$.
{Here, $\Delta_k$ is a free cumulant of ${\cal O}_{\Delta E}$,}
and $\lambda_k$  is an operator and system-specific non-universal coefficient.}
	Relations \eqref{eq-Deltak-DE} provide an infinite set of necessary conditions (signatures)  of emergent unitary symmetry. We confirm numerically this behavior emerges for all considered operators in a generic nonintegrable quantum spin systems, while also show that in the integrable case this behavior is not present.

	{In our numerical analysis, we identify $\Delta E_U$ with
the largest energy window for which \eqref{eq-Deltak-DE} holds for all
$\Delta_k$ that we are able to compute. More generally, one can define a {\it
cascade} of decreasing scales $\Delta E_U^{(\ell)} \geq \Delta E_U^{(\ell+1)}$,
where $\Delta E_U^{(\ell)}$ is defined as the maximal window for which
\eqref{eq-Deltak-DE} holds for all $k\leq \ell$. Thus, $\Delta E_U$ is
identified with $\Delta E_U^{(\ell)}$ for $\ell\rightarrow \infty$.}
	
	Emergent unitary symmetry has a clear manifestation at the level of matrix-element correlation functions, captured by the framework of general ETH  \cite{PhysRevLett.129.170603,Pappalardi:2023nsj,Fava:2023pac}. The latter proposes that averaged $k$-th cumulant  of ${ \cal O}_{mn}$ is given by a smooth function $f_k$ of $k$ energies $E_i$. Emergent unitary symmetry predicts that $f_k$ will be constant \cite{Pappalardi:2023nsj}, which will apply when all $E_i$ are within a narrow energy window of size $\Delta E_U^{(k)}$.
	For $k=2$ this is the condition  that function $f^2(\omega)\equiv f_2(\omega)$ entering the ETH ansatz for $m\neq n$,
	\begin{equation}
	\nonumber
	{\cal O}_{mn}=e^{-S/2}f(\omega) r_{mn},\,  \omega=(E_m-E_n), \, \overline{r^2}=1,
	\end{equation}
	for $|\omega|\leq \Delta E^{(2)}_U$ is constant.
	We can readily identify $\Delta E^{(2)}_U\equiv \Delta E_T$ with inverse thermalization timescale of $\cal O$, defined as the size of the short-frequency plateau of $f^2(\omega)$.
	We provide more technical details connecting our results with the framework of general ETH in Supplemental Material \cite{SuppMat}.

	For any $\Delta E\leq \Delta E_U$ unitary symmetry fixes the spectrum of ${\cal O}_{\Delta E}$, and the collective statistics of its matrix elements, in terms of the spectrum of ${\cal O}_U\equiv {\cal O}_{{\Delta E}_{U}}$. In particular, for $\Delta E\leq \Delta E_U$, ${\cal O}_{\Delta E}$ should in principle admit a description in terms of a rotational-invariant random-matrix model \cite{PhysRevLett.123.260601}, with the parameters of the model fine-tuned to match the spectrum of ${\cal O}_U$. We emphasize, matrix elements within the microcanonical window ${\Delta E}_U$  will show non-trivial correlations constrained only by unitary symmetry. 
	But when truncated to much more narrow windows $\Delta E\ll \Delta E_U$ these correlations will gradually disappear. 
	As follows from \eqref{eq-Deltak-DE}, for $k>2$, $\Delta_k/\Delta^{k/2}_2 \propto (\Delta E/\Delta E_U)^{(k/2 -1)} \rightarrow 0$, which implies generic random matrix model reduces to the Gaussian random matrix with uncorrelated ${\cal O}_{mn}$ and vanishing cumulants $\Delta_k$. 
	The onset of {\it Gaussian RMT}, and corresponding scale  
	$\Delta E_{GUE}$, were previously  scrutinized numerically in \cite{PhysRevLett.128.190601,PhysRevE.102.042127,PhysRevLett.128.180601}.

	Before embarking on concrete models and numerics, it is natural to ask how $\Delta E_U$ would depend on 
	the system size and the type of operator considered. Since the emergent unitary symmetry reshuffles energy {\it eigenstates}, it is tempting to identify $\Delta E_U$ with the Thouless energy scale $\Delta E_{Th}$ which marks the onset of random-matrix statistics of energy levels, i.e., emergent unitary symmetry reshuffling energy {\it eigenvalues}. 
	To complete  this proposal, we take into account  that Thouless energy controls thermalization time of slowest transport mode present in the system \cite{altshuler1986repulsion,PhysRevLett.123.210603,PhysRevX.12.021009,PhysRevB.105.104509,winer2023emergent,winer2023reappearance,Roy_Thouless20,Roy_Thouless22,Roy_Thouless23}. Hence, for a general operator  Thouless scale would coincide with  inverse thermalization timescale $\Delta E_T$. 
	Given that $\omega$-independence of $\overline{{ O}_{mn}^2}$ is a prerequisite for unitary symmetry, this proposal seems natural. {For a generic operator it ties Thouless energy with inverse thermalization time and the scale of unitary symmetry, $\Delta E_{Th}\sim \Delta E_T\equiv \Delta E_U^{(2)} \sim \Delta E_U=\Delta E_U^{(k)}$, $k\rightarrow \infty$. }

	This picture was first outlined in \cite{Review} but it can not be correct in general. As we discussed above, Gaussian random-matrix scale $\Delta E_{GUE}$ is expected to be smaller, but {\it not  parametrically} smaller than $\Delta E_U$. At the same time, in one-dimensional systems of length $L$, $\Delta E_{GUE}$ has to be {\it parametrically} smaller than $\Delta E_T$, specifically $\Delta E_{GUE}\propto \Delta E_T/L$ \cite{PhysRevLett.128.190601}, which contradicts $\Delta E_U\sim \Delta E_T$. 
	This contradiction is further elaborated in Supplemental Material~\cite{SuppMat}, where we provide accurate definitions for all scales. 
	{The important point is the cascade of scales 
		\begin{equation}
		\nonumber
		\Delta E_{Th}\sim \Delta E_U^{(2)} \geq \dots \geq \Delta E_U^{(k)} \geq \dots \geq \Delta E_U \geq \Delta E_{GUE}, 
		\end{equation}
		with the {\it parametric} difference between $\Delta E_{Th}\sim \Delta E_U^{(2)}$ and $\Delta E_U$.}
	We leave the question of systematically understanding system-size dependence of $\Delta E_U^{(k)}$ for the future, but note that $\Delta E_U$ evaluated numerically below vary significantly for different operators.


	{\it Models and Observables. }
	We proceed to study $\Delta_{k}(\Delta E)$ numerically in a chaotic many-body quantum system, where we consider a one-dimensional Ising model with transverse and longitudinal 
	fields, 
	\begin{equation}\label{eq-HIsing}
	H = \sum_{\ell = 1}^L \left(g \sigma_x^\ell+ h \sigma_z^\ell + J 
	\sigma_z^\ell \sigma_z^{\ell+1} \right)\ .
	\end{equation}
	$\sigma_{x,z}^\ell$ are Pauli spin operators at site $\ell$ and periodic 
	boundary conditions are employed, $\sigma_{x,z}^{L+1} \equiv \sigma_{x,z}^1$. 
	We set $J = 1$ and choose the fields as $g = 1$ and $h = 0.5$ for 
	which $H$ is 
	chaotic and expected to fulfill the ETH \cite{Banuls2011, RodriguezNieva2023}. To break 
	the translational and reflection symmetries of $H$, we further 
	add to $H$ two defect terms 
	$h_{d}\sigma_{z}^{\lfloor\frac{L}{3}\rfloor}$ and $-h_d\sigma_{z}^{\lfloor\frac{2L}{3}\rfloor}$ with $h_d = 0.02775$. Our numerical simulations are thus performed in the full Hilbert 
	space of dimension $D = 2^L$.  
	\begin{figure}[tb]
		\centering
		\includegraphics[width=1.0\linewidth]{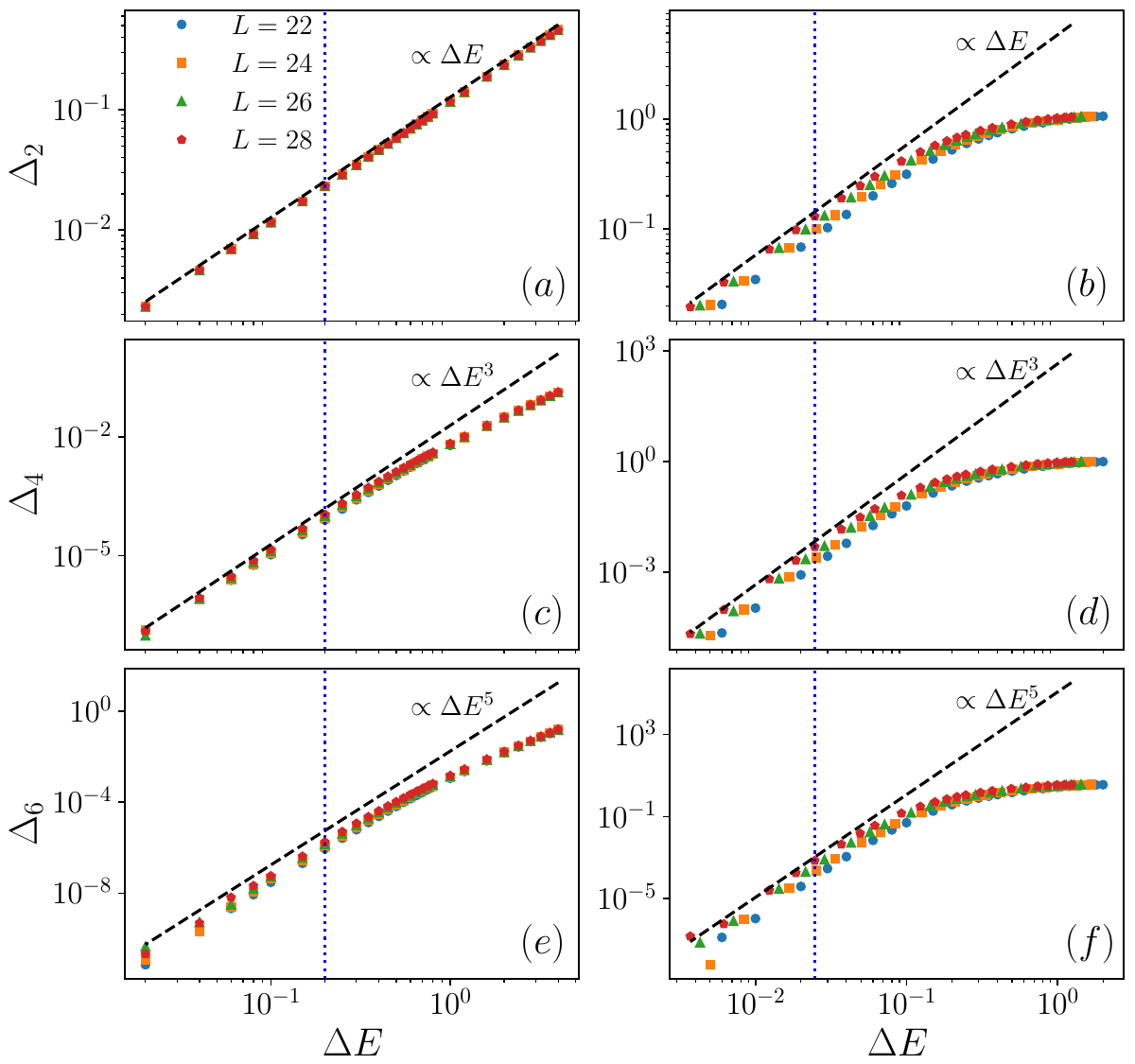}
		\caption{Even cumulants $\Delta_k$ for $k=2,4,6$ as a function of
$\Delta E$ for the density-wave operator ${\cal A}_q$ with the wave-number
$q={L}/{2}$ (panels  [(a),(c),(e)])  and with  $q = 1$ (panels [(b),(d),(f)]).
{Data are shown for different system sizes $L=22,24,26, 28$.
				As a guide to the eye, the inclined dashed lines (black) and
vertical dotted lines (blue) indicate the theoretically predicted slope
$\ln\Delta_{k}=(k-1)\ln \Delta E+{\rm const}$, and  an approximate location of
$\Delta E_U$, respectively.}
		}\label{Fig-DW}
	\end{figure}
	We study density-wave operators of the form,
	\begin{equation}{O}_q = \frac{1}{\sqrt{L}}\sum_{\ell=1}^L 
	\cos\left(\frac{2\pi}{L}q\ell\right){O}^\ell\ ,
	\end{equation}
	where we consider different momenta $q = 0,1,L/2$ and two different local $O^\ell$, i.e.,
	\begin{align}
	{\cal A}^{\ell} & =\frac{h}{2}(\sigma_{x}^{\ell}+\sigma_{x}^{\ell+1})+\frac{g}{2}(\sigma_{z}^{\ell}+\sigma_{z}^{\ell+1})+J\sigma_{z}^{\ell}\sigma_{z}^{\ell+1}\ , \nonumber \\
	{\cal B}^\ell & = \frac{
		g\sigma_z^\ell - h\sigma_x^\ell}{\sqrt{g^2 + h^2}}\ , 
	\end{align}
	with ${\cal A}^\ell$ being the energy density and 
	${\cal B}^\ell$ being constructed to have a small overlap with $H$ \cite{PhysRevE.99.010102}. Both operators behave in  agreement with the usual 
	indicators of the ETH \cite{PhysRevE.99.010102, PhysRevLett.128.180601} and we denote the corresponding density-wave operators by ${\cal A}_q$ and ${\cal B}_q$. Numerical data for the local operators ${\cal A}^\ell,{\cal B}^\ell$ can be found in \cite{SuppMat}.
	\begin{figure}[tb]
		\includegraphics[width=1.0\linewidth]{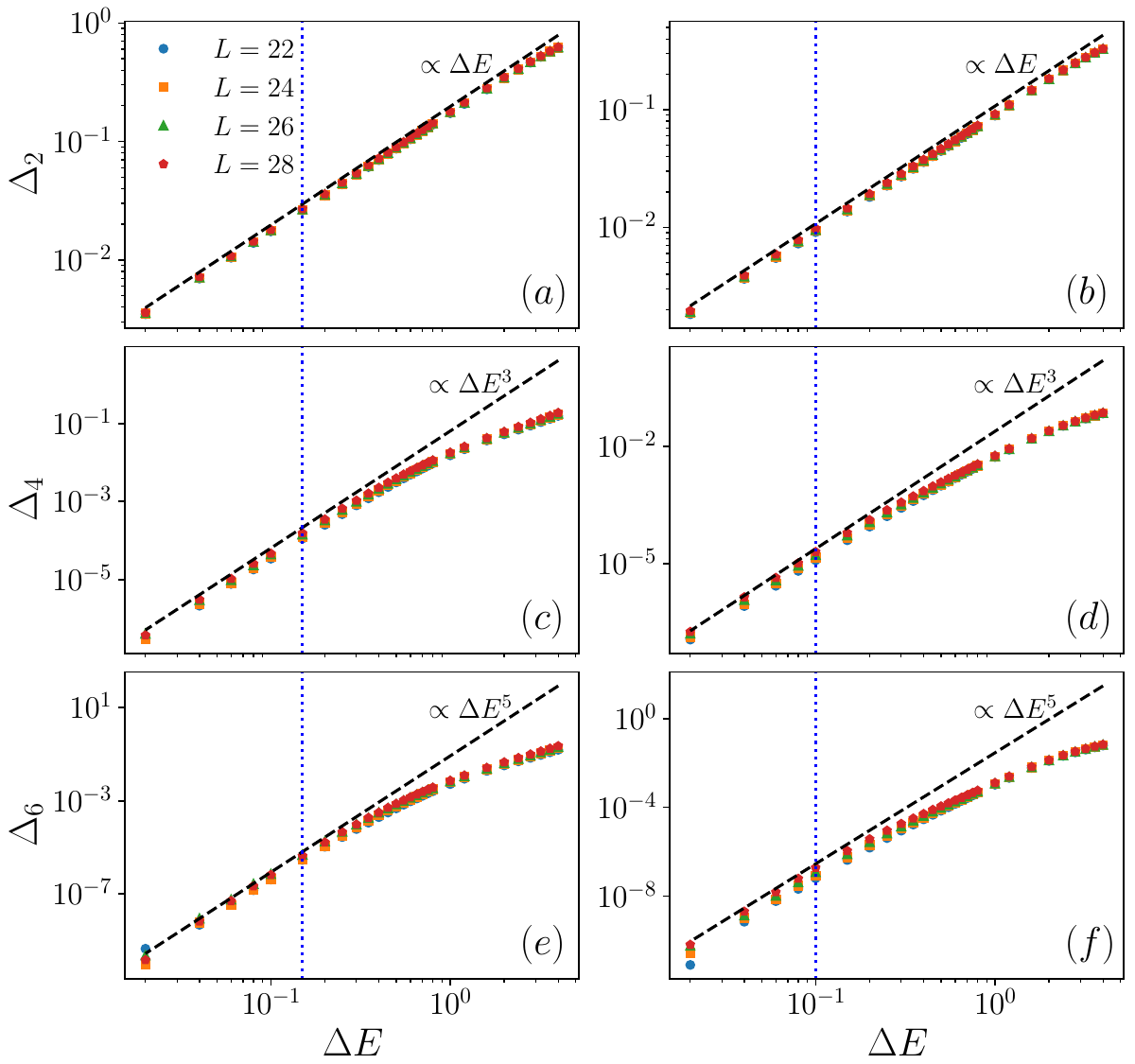}
		\caption{Similar to Fig. \ref{Fig-DW}, the results for density-wave operator ${\cal B}_q$ with $q=\frac{L}{2}$ (panels [(a),(c),(e)]) and $q = 1$ (panels [(b),(d),(f)]).}\label{Fig-Bq}
	\end{figure}
	%
	
	{\it Numerical approach. }
	{While studying the properties of ${\cal O}_{\Delta E}$  \cite{diag} would normally require full exact diagonalization (ED), we  can  go beyond system sizes accessible to standard ED to compute the moments ${\cal M}_k(\Delta E)$ \eqref{Eq::Mom}.}

	To evaluate moments, we exploit a pure-state technique based on the concept of quantum 
	typicality \cite{Heitmann2020,Jin2021} to compute the moments of ${\cal 
		O}_{\Delta E}$ (see also Refs.\ \cite{PhysRevLett.128.180601} and \cite{SuppMat} for more details).  
	A key idea within this approach is to realize that the energy filter 
	$P_{\Delta E}$ in Eq.\ \eqref{Eq::EFilter} can be expanded as \cite{PhysRevLett.128.180601}, 
	$P_{\Delta E} \ket{\psi} \simeq  \sum_{i=0}^{N_\text{tr}} C_i T_i(\tfrac{H-b}{a})\ket{\psi}$, where $T_i(x)$ are Chebyshev polynomials of the first kind, $C_i$ are 
	appropriately chosen coefficients that encode the energy window $\Delta E$ (see 
	\cite{PhysRevLett.128.180601, SuppMat}), and $a = 
	(E_\text{max}-E_\text{min})/2$, $b = (E_\text{max} + E_\text{min})/2$, where 
	$E_\text{max(min)}$ are the extremal eigenvalues of $H$. Moreover, the 
	expansion order $N_\text{tr}$ has to be chosen large enough to yield accurate results. 
	Given $P_{\Delta E}$, the trace in Eq.\ 
	\eqref{Eq::Mom} can 
	then be approximated by expectation values with respect 
	to random pure states, e.g., for ${\cal M}_k$ we have
	${\cal M}_{k}\approx\frac{\langle\psi|(P_{\Delta E}{\cal O}P_{\Delta E})^{k}|\psi\rangle}{\langle\psi|P_{\Delta E}|\psi\rangle}$ , 
	\
	with $\ket{\psi}$ being a Haar-random state constructed in the computational 
	basis. According to quantum typicality, the accuracy of this approximation 
	improves
	with the number of states $d$ inside the energy window. The accuracy can be further improved by averaging over different realization of random states.
	By applying $P_{\Delta E}$ efficiently using sparse matrices, 
	we are able to study the free cumulants $\Delta_k$ 
	for systems up to $L = 28$.

\begin{figure}[tb]
		\centering
		\includegraphics[width=1.0\linewidth]{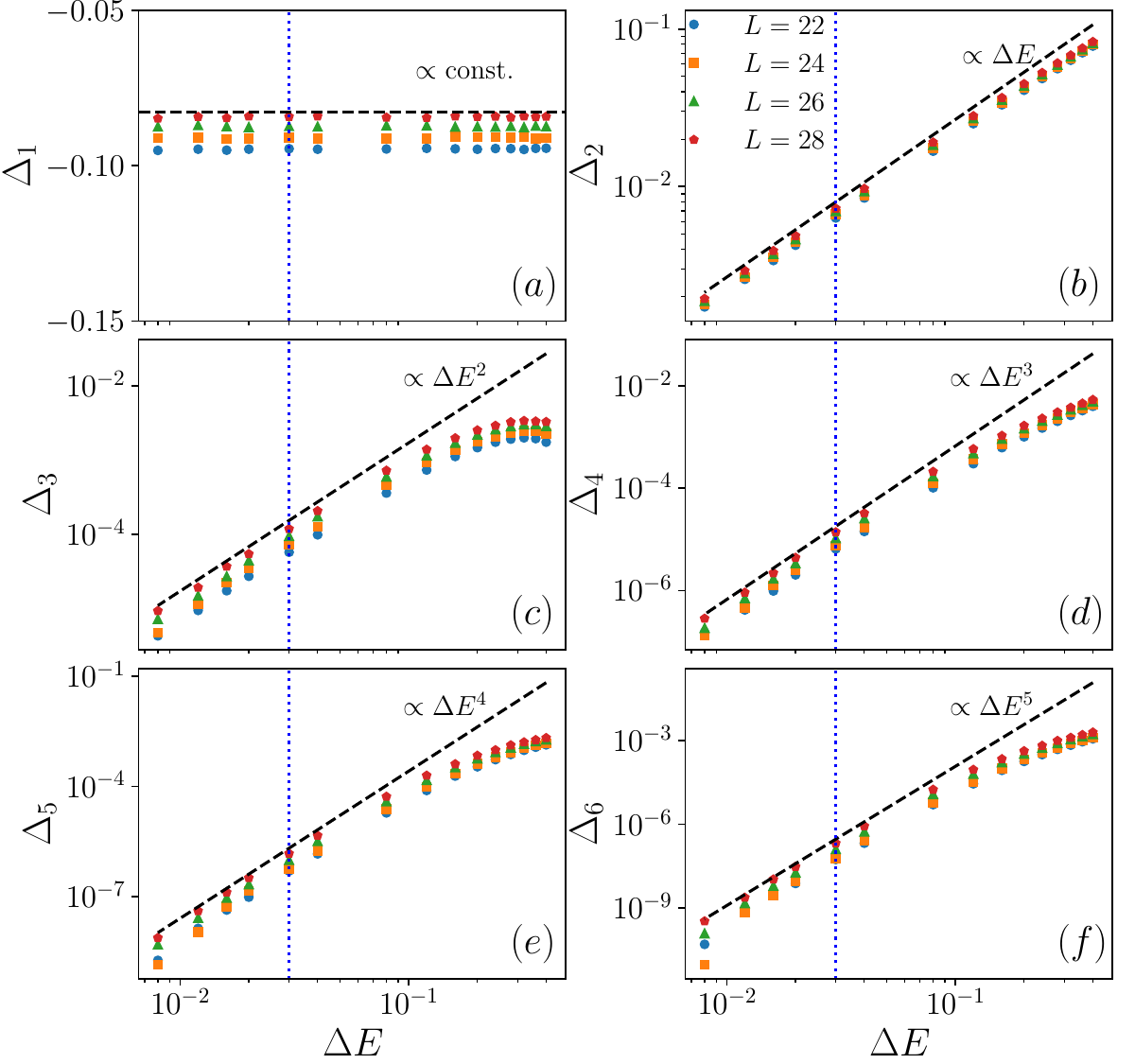}
		\caption{Similar to Figs.~\ref{Fig-DW}, the results for cumulants $\Delta_k$ for $k\leq 6$ as a function of $\Delta E$ for the density-wave operator ${\cal B}_q$ with $q=0$. 
		}\label{Fig-Bq0}
	\end{figure}

	{\it Results. }  
	In Fig.\ \ref{Fig-DW}, we show cumulants $\Delta_k$ as a function of energy window width $\Delta E$, for the energy density-wave operators ${\cal A}_q$ for two different wave numbers $q = L/2$ and $q = 1$.  
	Note that for the density-wave operators with $q > 0$ all odd cumulants approximately vanish (for $q = 0$, ${\cal A}_q$ is the Hamiltonian apart from the small defect), hence we only show results for even ones $\Delta_{2,4,6}$. 
	In Fig.\ \ref{Fig-DW}, we observe $\Delta_k \propto \Delta E^{k-1}$  behavior of Eq.~\eqref{eq-Deltak-DE} at sufficiently small energy scales, $\Delta E \le \Delta E_U$, indicating the onset of emergent unitary symmetry. The deviations from power-law behavior at extremely small $\Delta E$ are due to numerical errors aggravated by small number of states within such energy windows.
	A very similar picture emerges for the operators ${\cal B}_q$ with $q=L/2,1$ in Fig.\ \ref{Fig-Bq}, with the behavior \eqref{eq-Deltak-DE} clearly visible  below certain energy scale. We further observe the predicted power-law behavior for ${\cal B}_q$ with $q = 0$ in Fig.\ \ref{Fig-Bq0}, where we consider $k \leq 6$ including odd orders. Our results clearly  support emergence of unitary symmetry for all operators considered. 
	In the Supplemental material \cite{SuppMat}, we analyze local operators ${\cal A}^\ell,{\cal B}^\ell$ and  show existence of $\Delta E_U$, and expected power-law behavior below this scale, in these cases as well.

Combined with our theoretical results \eqref{eq-Deltak-DE},
Figs.\ \ref{Fig-DW} - \ref{Fig-Bq0} represent main contribution of this work. Our results demonstrate  existence of scale $\Delta E_U$ below which the statistical properties of matrix elements ${\cal O}_{mn}$ exhibit RMT universality while still admitting non-trivial correlations. Furthermore, scaling behavior \eqref{eq-Deltak-DE}, confirmed in Figs.\ \ref{Fig-DW} - \ref{Fig-Bq0}, implies $\Delta_{k}/(\Delta_2)^{k/2}\to 0$ when $\Delta E \to 0$ (up to statistical fluctuations), which confirms that ${\cal O}_{\Delta E}$ approaches a Gaussian random matrix at even smaller scales.
	
	Comparing results of  $\Delta_k$ for different $k$, 
	we find that in most cases, the size of the energy window marking the onset of $\Delta_k \propto (\Delta E)^{k-1}$ behavior is similar for higher cumulants with $k > 2$, but typically much smaller than the region of linear growth of $\Delta_2$
	(the operator ${\cal A}_1$ is an exception, see Fig.\ \ref{Fig-DW}~(b),(d),(f)). This difference between $\Delta E_U^{(2)}$ and $\Delta E_U^{(k)}$ for $k>2$ could be related to the fact that correlations between matrix elements have a fingerprint only in higher cumulants $\Delta_{k>2}$. 
	{Our data thus suggest that $\Delta E_U$ (vertical dashed line in Figs.\ \ref{Fig-DW} - \ref{Fig-Bq0}) in most cases is much smaller than inverse thermalization time $\Delta E_U^{(2)}\equiv \Delta E_T$, tentatively suggesting a scenario when in the thermodynamic limit 
	$\Delta E_U^{(2)}\ll \Delta E_U^{(3)} \sim \dots \sim \Delta E_U$.
	A conclusive analysis of $L$-dependence of $\Delta E_U^{(k)}$ would likely require system sizes much larger than numerically available. Available data in Figs.\ \ref{Fig-DW} - \ref{Fig-Bq0} (as well as additional data in the Supplemental Material \cite{SuppMat}) only indicates that $\Delta E_U$ decreases for larger system sizes.}
	

	{\it Conclusion \& Outlook. }
	In this paper we have introduced a novel energy scale $\Delta E_U$ marking the onset of emergent unitary symmetry, i.e., unitary random matrix theory universality of ${\cal O}_{mn}$ -- matrix elements  entering the eigenstate thermalization hypothesis.
	The scale is operator-specific. For $|E_n-E_m|\leq \Delta E_U$, matrix elements
	exhibit  unitary symmetry, which governs their collective statistical properties. {This is to say, below this scale ${ \cal O}_{mn}$ can be described in terms of a generic RMT. }
	To probe this scale we considered  a truncation of  an operator $\mathcal O$ to microcanonical energy window of size $\Delta E$, defined in \eqref{OdE}. We have shown that emergent unitary symmetry for $\Delta E\leq \Delta E_U$ is manifest through a simple power law $\Delta E$ dependence of free cumulants $\Delta_k$~\eqref{Eq::cumulants}, 
	\begin{eqnarray}
	\label{signature}
	\Delta_k(\Delta E)\propto (\Delta E)^{k-1}\ .
	\end{eqnarray}
	This provides a set of readily testable criteria, which can be accessed numerically beyond exact diagonalization. 
	
	We tested this behavior numerically for different operators satisfying ETH in the case of a generic quantum chaotic spin chain, and in each case found explicit evidence of \eqref{signature}. {Corresponding values of $\Delta E_U$ marking the onset of this behavior vary significantly for different operators, as well as the ratio $\Delta E_U/\Delta E_T$, where $\Delta E_T$ is operator's inverse thermalization time.} Our finite-size scaling analysis is not conclusive but consistent with theoretical expectation that in the large system size limit energy scales form the hierarchy 
	$\Delta E_{GUE} \leq \Delta E_{U} \ll \Delta E_T$, see \cite{SuppMat} for details.
	In particular when $L\rightarrow \infty$, for the considered one-dimensional system we expect both $\Delta E_{GUE}$ and $\Delta E_{U}$ to exhibit the same system-size dependence and be  parametrically smaller than $\Delta E_T$. 
	
	{Our results provide a unifying picture, connecting spectral properties of microcanonically-truncated operators \cite{PhysRevE.99.010102, PhysRevB.99.224302,PhysRevLett.128.190601,PhysRevE.102.042127,PhysRevLett.128.180601, pappalardi2023microcanonical,iniguez2023microcanonical} and correlations between off-diagonal matrix elements \cite{FoiniPRE,Murthy2019,Chan:2018fsp,Brenes2021,Hahn2023,PhysRevLett.123.260601,Belin:2021ryy,Anous:2021caj,Jafferis:2022uhu,Jafferis:2022wez,Belin:2023efa}.} An implicit question underlying these studies is to identify the degree of universality exhibited by the off-diagonal matrix elements of a typical operator in a generic quantum system. An analogous question about universality of energy spectrum of quantum chaotic systems is answered by the famous
	Bohigas-Giannoni-Schmit conjecture \cite{Bohigas}, which postulates random matrix universality (with global symmetries of the matrix model matching those of the original system). For the off-diagonal matrix elements, a conceptually similar proposal is given by a general random matrix theory.
	The question of universality was recently investigated  for holographic models of quantum gravity \cite{Jafferis:2022uhu,Jafferis:2022wez,Belin:2023efa}, yielding a multi-matrix model description specific for those cases. Our work, which studies a generic quantum chaotic model, provides evidence that emergent unitary symmetry is indeed the universal description for all operators satisfying ETH, with an operator-specific validity range $\Delta E_U$. This picture constitutes a compelling analog of the Bohigas-Giannoni-Schmit conjecture, and deserves further investigation. A natural question would be to repeat our analysis for different types of quantum chaotic systems, 
	including time-dependent Floquet models without energy conservation, as well as semi-classical few-body systems with a chaotic classical counterpart. We also emphasize the question of understanding relative scaling of  $\Delta E_U$, $\Delta E_T$ and $\Delta E_{Th}$ as a key to unite emergent unitary symmetry of this paper with the RMT universality of energy spectrum in one comprehensive framework of quantum chaos. \\
	\begin{acknowledgments}
	{\it Acknowledgements.}
	We thank Eugene Kanzieper for extensive discussions concerning random unitary projectors. This work has been funded by the Deutsche
	Forschungsgemeinschaft (DFG), under Grant No. 531128043, as well as under Grant
	No.\ 397107022, No.\ 397067869, and No.\ 397082825 within the DFG Research
	Unit FOR 2692, under Grant No.\ 355031190.
	A.\,D.\ is supported by the National Science Foundation under Grant No.~PHY 2310426. 
	This work was performed in part at Aspen Center for Physics, which is supported by National Science Foundation grant PHY-2210452.
	J.\,R.\ acknowledges funding from the European Union's Horizon Europe research 
	and innovation programme, 
	Marie Sk\l odowska-Curie grant no.\ 101060162, and the Packard Foundation
	through a Packard Fellowship in Science and Engineering.
	Additionally, we greatly acknowledge computing time on the HPC3 at the University of Osnabr\"{u}ck, granted by the DFG, under Grant No. 456666331.
	\end{acknowledgments}
	

	\bibliographystyle{apsrev4-1_titles}
	\bibliography{Ref.bib}

\begin{thebibliography}{49}%
\makeatletter
\providecommand \@ifxundefined [1]{%
 \@ifx{#1\undefined}
}%
\providecommand \@ifnum [1]{%
 \ifnum #1\expandafter \@firstoftwo
 \else \expandafter \@secondoftwo
 \fi
}%
\providecommand \@ifx [1]{%
 \ifx #1\expandafter \@firstoftwo
 \else \expandafter \@secondoftwo
 \fi
}%
\providecommand \natexlab [1]{#1}%
\providecommand \enquote  [1]{``#1''}%
\providecommand \bibnamefont  [1]{#1}%
\providecommand \bibfnamefont [1]{#1}%
\providecommand \citenamefont [1]{#1}%
\providecommand \href@noop [0]{\@secondoftwo}%
\providecommand \href [0]{\begingroup \@sanitize@url \@href}%
\providecommand \@href[1]{\@@startlink{#1}\@@href}%
\providecommand \@@href[1]{\endgroup#1\@@endlink}%
\providecommand \@sanitize@url [0]{\catcode `\\12\catcode `\$12\catcode
  `\&12\catcode `\#12\catcode `\^12\catcode `\_12\catcode `\%12\relax}%
\providecommand \@@startlink[1]{}%
\providecommand \@@endlink[0]{}%
\providecommand \url  [0]{\begingroup\@sanitize@url \@url }%
\providecommand \@url [1]{\endgroup\@href {#1}{\urlprefix }}%
\providecommand \urlprefix  [0]{URL }%
\providecommand \Eprint [0]{\href }%
\providecommand \doibase [0]{http://dx.doi.org/}%
\providecommand \selectlanguage [0]{\@gobble}%
\providecommand \bibinfo  [0]{\@secondoftwo}%
\providecommand \bibfield  [0]{\@secondoftwo}%
\providecommand \translation [1]{[#1]}%
\providecommand \BibitemOpen [0]{}%
\providecommand \bibitemStop [0]{}%
\providecommand \bibitemNoStop [0]{.\EOS\space}%
\providecommand \EOS [0]{\spacefactor3000\relax}%
\providecommand \BibitemShut  [1]{\csname bibitem#1\endcsname}%
\let\auto@bib@innerbib\@empty
\bibitem [{\citenamefont {Deutsch}(1991)}]{Deutsch}%
  \BibitemOpen
  \bibfield  {author} {\bibinfo {author} {\bibfnamefont {J.~M.}\ \bibnamefont
  {Deutsch}},\ }\bibfield  {title} {\emph {\bibinfo {title} {Quantum
  statistical mechanics in a closed system},\ }}\href {\doibase
  10.1103/PhysRevA.43.2046} {\bibfield  {journal} {\bibinfo  {journal} {Phys.
  Rev. A}\ }\textbf {\bibinfo {volume} {43}},\ \bibinfo {pages} {2046}
  (\bibinfo {year} {1991})}\BibitemShut {NoStop}%
\bibitem [{\citenamefont {Srednicki}(1994)}]{Srednicki94}%
  \BibitemOpen
  \bibfield  {author} {\bibinfo {author} {\bibfnamefont {M.}~\bibnamefont
  {Srednicki}},\ }\bibfield  {title} {\emph {\bibinfo {title} {Chaos and
  quantum thermalization},\ }}\href {\doibase 10.1103/PhysRevE.50.888}
  {\bibfield  {journal} {\bibinfo  {journal} {Phys. Rev. E}\ }\textbf {\bibinfo
  {volume} {50}},\ \bibinfo {pages} {888} (\bibinfo {year} {1994})}\BibitemShut
  {NoStop}%
\bibitem [{\citenamefont {Luca~D'Alessio}\ and\ \citenamefont
  {Rigol}(2016)}]{Review}%
  \BibitemOpen
  \bibfield  {author} {\bibinfo {author} {\bibfnamefont {A.~P.}\ \bibnamefont
  {Luca~D'Alessio}, \bibfnamefont {Yariv~Kafri}}\ and\ \bibinfo {author}
  {\bibfnamefont {M.}~\bibnamefont {Rigol}},\ }\bibfield  {title} {\emph
  {\bibinfo {title} {From quantum chaos and eigenstate thermalization to
  statistical mechanics and thermodynamics},\ }}\href {\doibase
  10.1080/00018732.2016.1198134} {\bibfield  {journal} {\bibinfo  {journal}
  {Advances in Physics}\ }\textbf {\bibinfo {volume} {65}},\ \bibinfo {pages}
  {239} (\bibinfo {year} {2016})}\BibitemShut {NoStop}%
\bibitem [{\citenamefont {Dymarsky}\ \emph {et~al.}(2018)\citenamefont
  {Dymarsky}, \citenamefont {Lashkari},\ and\ \citenamefont
  {Liu}}]{PhysRevE.97.012140}%
  \BibitemOpen
  \bibfield  {author} {\bibinfo {author} {\bibfnamefont {A.}~\bibnamefont
  {Dymarsky}}, \bibinfo {author} {\bibfnamefont {N.}~\bibnamefont {Lashkari}},
  \ and\ \bibinfo {author} {\bibfnamefont {H.}~\bibnamefont {Liu}},\ }\bibfield
   {title} {\emph {\bibinfo {title} {Subsystem eigenstate thermalization
  hypothesis},\ }}\href {\doibase 10.1103/PhysRevE.97.012140} {\bibfield
  {journal} {\bibinfo  {journal} {Phys. Rev. E}\ }\textbf {\bibinfo {volume}
  {97}},\ \bibinfo {pages} {012140} (\bibinfo {year} {2018})}\BibitemShut
  {NoStop}%
\bibitem [{\citenamefont {Pappalardi}\ \emph {et~al.}(2022)\citenamefont
  {Pappalardi}, \citenamefont {Foini},\ and\ \citenamefont
  {Kurchan}}]{PhysRevLett.129.170603}%
  \BibitemOpen
  \bibfield  {author} {\bibinfo {author} {\bibfnamefont {S.}~\bibnamefont
  {Pappalardi}}, \bibinfo {author} {\bibfnamefont {L.}~\bibnamefont {Foini}}, \
  and\ \bibinfo {author} {\bibfnamefont {J.}~\bibnamefont {Kurchan}},\
  }\bibfield  {title} {\emph {\bibinfo {title} {Eigenstate thermalization
  hypothesis and free probability},\ }}\href {\doibase
  10.1103/PhysRevLett.129.170603} {\bibfield  {journal} {\bibinfo  {journal}
  {Phys. Rev. Lett.}\ }\textbf {\bibinfo {volume} {129}},\ \bibinfo {pages}
  {170603} (\bibinfo {year} {2022})}\BibitemShut {NoStop}%
\bibitem [{\citenamefont {Bohigas}\ \emph {et~al.}(1984)\citenamefont
  {Bohigas}, \citenamefont {Giannoni},\ and\ \citenamefont {Schmit}}]{Bohigas}%
  \BibitemOpen
  \bibfield  {author} {\bibinfo {author} {\bibfnamefont {O.}~\bibnamefont
  {Bohigas}}, \bibinfo {author} {\bibfnamefont {M.-J.}\ \bibnamefont
  {Giannoni}}, \ and\ \bibinfo {author} {\bibfnamefont {C.}~\bibnamefont
  {Schmit}},\ }\bibfield  {title} {\emph {\bibinfo {title} {Characterization of
  chaotic quantum spectra and universality of level fluctuation laws},\
  }}\href@noop {} {\bibfield  {journal} {\bibinfo  {journal} {Physical review
  letters}\ }\textbf {\bibinfo {volume} {52}},\ \bibinfo {pages} {1} (\bibinfo
  {year} {1984})}\BibitemShut {NoStop}%
\bibitem [{\citenamefont {Richter}\ \emph {et~al.}(2020)\citenamefont
  {Richter}, \citenamefont {Dymarsky}, \citenamefont {Steinigeweg},\ and\
  \citenamefont {Gemmer}}]{PhysRevE.102.042127}%
  \BibitemOpen
  \bibfield  {author} {\bibinfo {author} {\bibfnamefont {J.}~\bibnamefont
  {Richter}}, \bibinfo {author} {\bibfnamefont {A.}~\bibnamefont {Dymarsky}},
  \bibinfo {author} {\bibfnamefont {R.}~\bibnamefont {Steinigeweg}}, \ and\
  \bibinfo {author} {\bibfnamefont {J.}~\bibnamefont {Gemmer}},\ }\bibfield
  {title} {\emph {\bibinfo {title} {Eigenstate thermalization hypothesis beyond
  standard indicators: Emergence of random-matrix behavior at small
  frequencies},\ }}\href {\doibase 10.1103/PhysRevE.102.042127} {\bibfield
  {journal} {\bibinfo  {journal} {Phys. Rev. E}\ }\textbf {\bibinfo {volume}
  {102}},\ \bibinfo {pages} {042127} (\bibinfo {year} {2020})}\BibitemShut
  {NoStop}%
\bibitem [{\citenamefont {Wang}\ \emph {et~al.}(2022)\citenamefont {Wang},
  \citenamefont {Lamann}, \citenamefont {Richter}, \citenamefont {Steinigeweg},
  \citenamefont {Dymarsky},\ and\ \citenamefont
  {Gemmer}}]{PhysRevLett.128.180601}%
  \BibitemOpen
  \bibfield  {author} {\bibinfo {author} {\bibfnamefont {J.}~\bibnamefont
  {Wang}}, \bibinfo {author} {\bibfnamefont {M.~H.}\ \bibnamefont {Lamann}},
  \bibinfo {author} {\bibfnamefont {J.}~\bibnamefont {Richter}}, \bibinfo
  {author} {\bibfnamefont {R.}~\bibnamefont {Steinigeweg}}, \bibinfo {author}
  {\bibfnamefont {A.}~\bibnamefont {Dymarsky}}, \ and\ \bibinfo {author}
  {\bibfnamefont {J.}~\bibnamefont {Gemmer}},\ }\bibfield  {title} {\emph
  {\bibinfo {title} {Eigenstate thermalization hypothesis and its deviations
  from random-matrix theory beyond the thermalization time},\ }}\href {\doibase
  10.1103/PhysRevLett.128.180601} {\bibfield  {journal} {\bibinfo  {journal}
  {Phys. Rev. Lett.}\ }\textbf {\bibinfo {volume} {128}},\ \bibinfo {pages}
  {180601} (\bibinfo {year} {2022})}\BibitemShut {NoStop}%
\bibitem [{\citenamefont {Dymarsky}(2022)}]{PhysRevLett.128.190601}%
  \BibitemOpen
  \bibfield  {author} {\bibinfo {author} {\bibfnamefont {A.}~\bibnamefont
  {Dymarsky}},\ }\bibfield  {title} {\emph {\bibinfo {title} {Bound on
  eigenstate thermalization from transport},\ }}\href {\doibase
  10.1103/PhysRevLett.128.190601} {\bibfield  {journal} {\bibinfo  {journal}
  {Phys. Rev. Lett.}\ }\textbf {\bibinfo {volume} {128}},\ \bibinfo {pages}
  {190601} (\bibinfo {year} {2022})}\BibitemShut {NoStop}%
\bibitem [{\citenamefont {Foini}\ and\ \citenamefont
  {Kurchan}(2019{\natexlab{a}})}]{FoiniPRE}%
  \BibitemOpen
  \bibfield  {author} {\bibinfo {author} {\bibfnamefont {L.}~\bibnamefont
  {Foini}}\ and\ \bibinfo {author} {\bibfnamefont {J.}~\bibnamefont
  {Kurchan}},\ }\bibfield  {title} {\emph {\bibinfo {title} {Eigenstate
  thermalization hypothesis and out of time order correlators},\ }}\href
  {\doibase 10.1103/PhysRevE.99.042139} {\bibfield  {journal} {\bibinfo
  {journal} {Phys. Rev. E}\ }\textbf {\bibinfo {volume} {99}},\ \bibinfo
  {pages} {042139} (\bibinfo {year} {2019}{\natexlab{a}})}\BibitemShut
  {NoStop}%
\bibitem [{\citenamefont {Chan}\ \emph {et~al.}(2019)\citenamefont {Chan},
  \citenamefont {De~Luca},\ and\ \citenamefont {Chalker}}]{Chan:2018fsp}%
  \BibitemOpen
  \bibfield  {author} {\bibinfo {author} {\bibfnamefont {A.}~\bibnamefont
  {Chan}}, \bibinfo {author} {\bibfnamefont {A.}~\bibnamefont {De~Luca}}, \
  and\ \bibinfo {author} {\bibfnamefont {J.~T.}\ \bibnamefont {Chalker}},\
  }\bibfield  {title} {\emph {\bibinfo {title} {Eigenstate correlations,
  thermalization, and the butterfly effect},\ }}\href {\doibase
  10.1103/PhysRevLett.122.220601} {\bibfield  {journal} {\bibinfo  {journal}
  {Phys. Rev. Lett.}\ }\textbf {\bibinfo {volume} {122}},\ \bibinfo {pages}
  {220601} (\bibinfo {year} {2019})}\BibitemShut {NoStop}%
\bibitem [{\citenamefont {Murthy}\ and\ \citenamefont
  {Srednicki}(2019)}]{Murthy2019}%
  \BibitemOpen
  \bibfield  {author} {\bibinfo {author} {\bibfnamefont {C.}~\bibnamefont
  {Murthy}}\ and\ \bibinfo {author} {\bibfnamefont {M.}~\bibnamefont
  {Srednicki}},\ }\bibfield  {title} {\emph {\bibinfo {title} {Bounds on chaos
  from the eigenstate thermalization hypothesis},\ }}\href {\doibase
  10.1103/PhysRevLett.123.230606} {\bibfield  {journal} {\bibinfo  {journal}
  {Phys. Rev. Lett.}\ }\textbf {\bibinfo {volume} {123}},\ \bibinfo {pages}
  {230606} (\bibinfo {year} {2019})}\BibitemShut {NoStop}%
\bibitem [{\citenamefont {Brenes}\ \emph {et~al.}(2021)\citenamefont {Brenes},
  \citenamefont {Pappalardi}, \citenamefont {Mitchison}, \citenamefont
  {Goold},\ and\ \citenamefont {Silva}}]{Brenes2021}%
  \BibitemOpen
  \bibfield  {author} {\bibinfo {author} {\bibfnamefont {M.}~\bibnamefont
  {Brenes}}, \bibinfo {author} {\bibfnamefont {S.}~\bibnamefont {Pappalardi}},
  \bibinfo {author} {\bibfnamefont {M.~T.}\ \bibnamefont {Mitchison}}, \bibinfo
  {author} {\bibfnamefont {J.}~\bibnamefont {Goold}}, \ and\ \bibinfo {author}
  {\bibfnamefont {A.}~\bibnamefont {Silva}},\ }\bibfield  {title} {\emph
  {\bibinfo {title} {Out-of-time-order correlations and the fine structure of
  eigenstate thermalization},\ }}\href {\doibase 10.1103/PhysRevE.104.034120}
  {\bibfield  {journal} {\bibinfo  {journal} {Phys. Rev. E}\ }\textbf {\bibinfo
  {volume} {104}},\ \bibinfo {pages} {034120} (\bibinfo {year}
  {2021})}\BibitemShut {NoStop}%
\bibitem [{\citenamefont {Foini}\ and\ \citenamefont
  {Kurchan}(2019{\natexlab{b}})}]{PhysRevLett.123.260601}%
  \BibitemOpen
  \bibfield  {author} {\bibinfo {author} {\bibfnamefont {L.}~\bibnamefont
  {Foini}}\ and\ \bibinfo {author} {\bibfnamefont {J.}~\bibnamefont
  {Kurchan}},\ }\bibfield  {title} {\emph {\bibinfo {title} {Eigenstate
  thermalization and rotational invariance in ergodic quantum systems},\
  }}\href {\doibase 10.1103/PhysRevLett.123.260601} {\bibfield  {journal}
  {\bibinfo  {journal} {Phys. Rev. Lett.}\ }\textbf {\bibinfo {volume} {123}},\
  \bibinfo {pages} {260601} (\bibinfo {year} {2019}{\natexlab{b}})}\BibitemShut
  {NoStop}%
\bibitem [{Sup()}]{SuppMat}%
  \BibitemOpen
  \href@noop {} {}\bibinfo {note} {See supplemental material for the derivation
  of Eq.~\eqref{eq-Deltak-DE}, definition of $\lambda_k$, additional numerical
  results, rigorous definition of different energy scales, {discussions on
  unitary symmetry in integrable case} and details on our numerical approach.
  Supplemental material includes Refs.\
  \cite{forrester2006quantum,savin2008nonlinear,vidal2012statistics,Pucha_a_2017,
  Fehske06,Osborn:1998nf}.}\BibitemShut {Stop}%
\bibitem [{Sym()}]{Symmetry}%
  \BibitemOpen
  \href@noop {} {}\bibinfo {note} {{Unitary symmetry can be understood as
  rotational invariance of the probablity measure $P({\cal O}_{mn})=P({\cal
  O}'_{mn})$, ${\cal O}'=U{\cal O}U^\dagger$. The type of symmetry depends on
  global symmetries of $H$ and $O$. When both exhibit time-reversal symmetry,
  {\it unitary} symmetry $U(d)$ should be substituted by {\it orthogonal}
  symmetry. Nevertheless we use the notations $\Delta E_U$ (and $\Delta
  E_{GUE}$ for corresponding Gaussian ensemble) throughout the paper for the
  scales marking the onset of corresponding regimes.}}\BibitemShut {Stop}%
\bibitem [{\citenamefont {Dymarsky}\ and\ \citenamefont
  {Liu}(2019)}]{PhysRevE.99.010102}%
  \BibitemOpen
  \bibfield  {author} {\bibinfo {author} {\bibfnamefont {A.}~\bibnamefont
  {Dymarsky}}\ and\ \bibinfo {author} {\bibfnamefont {H.}~\bibnamefont {Liu}},\
  }\bibfield  {title} {\emph {\bibinfo {title} {New characteristic of quantum
  many-body chaotic systems},\ }}\href {\doibase 10.1103/PhysRevE.99.010102}
  {\bibfield  {journal} {\bibinfo  {journal} {Phys. Rev. E}\ }\textbf {\bibinfo
  {volume} {99}},\ \bibinfo {pages} {010102} (\bibinfo {year}
  {2019})}\BibitemShut {NoStop}%
\bibitem [{\citenamefont {Dymarsky}(2019)}]{PhysRevB.99.224302}%
  \BibitemOpen
  \bibfield  {author} {\bibinfo {author} {\bibfnamefont {A.}~\bibnamefont
  {Dymarsky}},\ }\bibfield  {title} {\emph {\bibinfo {title} {Mechanism of
  macroscopic equilibration of isolated quantum systems},\ }}\href {\doibase
  10.1103/PhysRevB.99.224302} {\bibfield  {journal} {\bibinfo  {journal} {Phys.
  Rev. B}\ }\textbf {\bibinfo {volume} {99}},\ \bibinfo {pages} {224302}
  (\bibinfo {year} {2019})}\BibitemShut {NoStop}%
\bibitem [{\citenamefont {Pappalardi}\ \emph {et~al.}(2024)\citenamefont
  {Pappalardi}, \citenamefont {Foini},\ and\ \citenamefont
  {Kurchan}}]{pappalardi2023microcanonical}%
  \BibitemOpen
  \bibfield  {author} {\bibinfo {author} {\bibfnamefont {S.}~\bibnamefont
  {Pappalardi}}, \bibinfo {author} {\bibfnamefont {L.}~\bibnamefont {Foini}}, \
  and\ \bibinfo {author} {\bibfnamefont {J.}~\bibnamefont {Kurchan}},\
  }\bibfield  {title} {\emph {\bibinfo {title} {Microcanonical windows on
  quantum operators},\ }}\href {\doibase 10.22331/q-2024-01-11-1227} {\bibfield
   {journal} {\bibinfo  {journal} {{Quantum}}\ }\textbf {\bibinfo {volume}
  {8}},\ \bibinfo {pages} {1227} (\bibinfo {year} {2024})}\BibitemShut
  {NoStop}%
\bibitem [{\citenamefont {Iniguez}\ and\ \citenamefont
  {Srednicki}(2023)}]{iniguez2023microcanonical}%
  \BibitemOpen
  \bibfield  {author} {\bibinfo {author} {\bibfnamefont {F.}~\bibnamefont
  {Iniguez}}\ and\ \bibinfo {author} {\bibfnamefont {M.}~\bibnamefont
  {Srednicki}},\ }\bibfield  {title} {\emph {\bibinfo {title} {Microcanonical
  truncations of observables in quantum chaotic systems},\ }}\href@noop {}
  {\bibfield  {journal} {\bibinfo  {journal} {arXiv preprint arXiv:2305.15702}\
  } (\bibinfo {year} {2023})}\BibitemShut {NoStop}%
\bibitem [{\citenamefont {Pappalardi}\ \emph {et~al.}(2023)\citenamefont
  {Pappalardi}, \citenamefont {Fritzsch},\ and\ \citenamefont
  {Prosen}}]{Pappalardi:2023nsj}%
  \BibitemOpen
  \bibfield  {author} {\bibinfo {author} {\bibfnamefont {S.}~\bibnamefont
  {Pappalardi}}, \bibinfo {author} {\bibfnamefont {F.}~\bibnamefont
  {Fritzsch}}, \ and\ \bibinfo {author} {\bibfnamefont {T.}~\bibnamefont
  {Prosen}},\ }\bibfield  {title} {\emph {\bibinfo {title} {General eigenstate
  thermalization via free cumulants in quantum lattice systems},\ }}\href@noop
  {} {\bibfield  {journal} {\bibinfo  {journal} {arXiv preprint
  arXiv:2303.00713}\ } (\bibinfo {year} {2023})}\BibitemShut {NoStop}%
\bibitem [{\citenamefont {Fava}\ \emph {et~al.}(2023)\citenamefont {Fava},
  \citenamefont {Kurchan},\ and\ \citenamefont {Pappalardi}}]{Fava:2023pac}%
  \BibitemOpen
  \bibfield  {author} {\bibinfo {author} {\bibfnamefont {M.}~\bibnamefont
  {Fava}}, \bibinfo {author} {\bibfnamefont {J.}~\bibnamefont {Kurchan}}, \
  and\ \bibinfo {author} {\bibfnamefont {S.}~\bibnamefont {Pappalardi}},\
  }\bibfield  {title} {\emph {\bibinfo {title} {Designs via free probability},\
  }}\href@noop {} {\bibfield  {journal} {\bibinfo  {journal} {arXiv preprint
  arXiv:2308.06200}\ } (\bibinfo {year} {2023})}\BibitemShut {NoStop}%
\bibitem [{\citenamefont {Altshuler}\ and\ \citenamefont
  {Shklovskii}(1986)}]{altshuler1986repulsion}%
  \BibitemOpen
  \bibfield  {author} {\bibinfo {author} {\bibfnamefont {B.}~\bibnamefont
  {Altshuler}}\ and\ \bibinfo {author} {\bibfnamefont {B.}~\bibnamefont
  {Shklovskii}},\ }\bibfield  {title} {\emph {\bibinfo {title} {Repulsion of
  energy levels and conductivity of small metal samples},\ }}\href@noop {}
  {\bibfield  {journal} {\bibinfo  {journal} {Sov. Phys. JETP}\ }\textbf
  {\bibinfo {volume} {64}},\ \bibinfo {pages} {127} (\bibinfo {year}
  {1986})}\BibitemShut {NoStop}%
\bibitem [{\citenamefont {Friedman}\ \emph {et~al.}(2019)\citenamefont
  {Friedman}, \citenamefont {Chan}, \citenamefont {De~Luca},\ and\
  \citenamefont {Chalker}}]{PhysRevLett.123.210603}%
  \BibitemOpen
  \bibfield  {author} {\bibinfo {author} {\bibfnamefont {A.~J.}\ \bibnamefont
  {Friedman}}, \bibinfo {author} {\bibfnamefont {A.}~\bibnamefont {Chan}},
  \bibinfo {author} {\bibfnamefont {A.}~\bibnamefont {De~Luca}}, \ and\
  \bibinfo {author} {\bibfnamefont {J.~T.}\ \bibnamefont {Chalker}},\
  }\bibfield  {title} {\emph {\bibinfo {title} {Spectral statistics and
  many-body quantum chaos with conserved charge},\ }}\href {\doibase
  10.1103/PhysRevLett.123.210603} {\bibfield  {journal} {\bibinfo  {journal}
  {Phys. Rev. Lett.}\ }\textbf {\bibinfo {volume} {123}},\ \bibinfo {pages}
  {210603} (\bibinfo {year} {2019})}\BibitemShut {NoStop}%
\bibitem [{\citenamefont {Winer}\ and\ \citenamefont
  {Swingle}(2022{\natexlab{a}})}]{PhysRevX.12.021009}%
  \BibitemOpen
  \bibfield  {author} {\bibinfo {author} {\bibfnamefont {M.}~\bibnamefont
  {Winer}}\ and\ \bibinfo {author} {\bibfnamefont {B.}~\bibnamefont
  {Swingle}},\ }\bibfield  {title} {\emph {\bibinfo {title} {Hydrodynamic
  theory of the connected spectral form factor},\ }}\href {\doibase
  10.1103/PhysRevX.12.021009} {\bibfield  {journal} {\bibinfo  {journal} {Phys.
  Rev. X}\ }\textbf {\bibinfo {volume} {12}},\ \bibinfo {pages} {021009}
  (\bibinfo {year} {2022}{\natexlab{a}})}\BibitemShut {NoStop}%
\bibitem [{\citenamefont {Winer}\ and\ \citenamefont
  {Swingle}(2022{\natexlab{b}})}]{PhysRevB.105.104509}%
  \BibitemOpen
  \bibfield  {author} {\bibinfo {author} {\bibfnamefont {M.}~\bibnamefont
  {Winer}}\ and\ \bibinfo {author} {\bibfnamefont {B.}~\bibnamefont
  {Swingle}},\ }\bibfield  {title} {\emph {\bibinfo {title} {Spontaneous
  symmetry breaking, spectral statistics, and the ramp},\ }}\href {\doibase
  10.1103/PhysRevB.105.104509} {\bibfield  {journal} {\bibinfo  {journal}
  {Phys. Rev. B}\ }\textbf {\bibinfo {volume} {105}},\ \bibinfo {pages}
  {104509} (\bibinfo {year} {2022}{\natexlab{b}})}\BibitemShut {NoStop}%
\bibitem [{\citenamefont {Winer}\ and\ \citenamefont
  {Swingle}(2023{\natexlab{a}})}]{winer2023emergent}%
  \BibitemOpen
  \bibfield  {author} {\bibinfo {author} {\bibfnamefont {M.}~\bibnamefont
  {Winer}}\ and\ \bibinfo {author} {\bibfnamefont {B.}~\bibnamefont
  {Swingle}},\ }\bibfield  {title} {\emph {\bibinfo {title} {Emergent spectral
  form factors in sonic systems},\ }}\href {\doibase
  10.1103/PhysRevB.108.054523} {\bibfield  {journal} {\bibinfo  {journal}
  {Phys. Rev. B}\ }\textbf {\bibinfo {volume} {108}},\ \bibinfo {pages}
  {054523} (\bibinfo {year} {2023}{\natexlab{a}})}\BibitemShut {NoStop}%
\bibitem [{\citenamefont {Winer}\ and\ \citenamefont
  {Swingle}(2023{\natexlab{b}})}]{winer2023reappearance}%
  \BibitemOpen
  \bibfield  {author} {\bibinfo {author} {\bibfnamefont {M.}~\bibnamefont
  {Winer}}\ and\ \bibinfo {author} {\bibfnamefont {B.}~\bibnamefont
  {Swingle}},\ }\bibfield  {title} {\emph {\bibinfo {title} {Reappearance of
  thermalization dynamics in the late-time spectral form factor},\ }}\href@noop
  {} {\bibfield  {journal} {\bibinfo  {journal} {arXiv preprint
  arXiv:2307.14415}\ } (\bibinfo {year} {2023}{\natexlab{b}})}\BibitemShut
  {NoStop}%
\bibitem [{\citenamefont {Roy}\ and\ \citenamefont
  {Prosen}(2020)}]{Roy_Thouless20}%
  \BibitemOpen
  \bibfield  {author} {\bibinfo {author} {\bibfnamefont {D.}~\bibnamefont
  {Roy}}\ and\ \bibinfo {author} {\bibfnamefont {T.~c.~v.}\ \bibnamefont
  {Prosen}},\ }\bibfield  {title} {\emph {\bibinfo {title} {Random matrix
  spectral form factor in kicked interacting fermionic chains},\ }}\href
  {\doibase 10.1103/PhysRevE.102.060202} {\bibfield  {journal} {\bibinfo
  {journal} {Phys. Rev. E}\ }\textbf {\bibinfo {volume} {102}},\ \bibinfo
  {pages} {060202} (\bibinfo {year} {2020})}\BibitemShut {NoStop}%
\bibitem [{\citenamefont {Roy}\ \emph {et~al.}(2022)\citenamefont {Roy},
  \citenamefont {Mishra},\ and\ \citenamefont {Prosen}}]{Roy_Thouless22}%
  \BibitemOpen
  \bibfield  {author} {\bibinfo {author} {\bibfnamefont {D.}~\bibnamefont
  {Roy}}, \bibinfo {author} {\bibfnamefont {D.}~\bibnamefont {Mishra}}, \ and\
  \bibinfo {author} {\bibfnamefont {T.~c.~v.}\ \bibnamefont {Prosen}},\
  }\bibfield  {title} {\emph {\bibinfo {title} {Spectral form factor in a
  minimal bosonic model of many-body quantum chaos},\ }}\href {\doibase
  10.1103/PhysRevE.106.024208} {\bibfield  {journal} {\bibinfo  {journal}
  {Phys. Rev. E}\ }\textbf {\bibinfo {volume} {106}},\ \bibinfo {pages}
  {024208} (\bibinfo {year} {2022})}\BibitemShut {NoStop}%
\bibitem [{\citenamefont {Kumar}\ and\ \citenamefont
  {Roy}(2024)}]{Roy_Thouless23}%
  \BibitemOpen
  \bibfield  {author} {\bibinfo {author} {\bibfnamefont {V.}~\bibnamefont
  {Kumar}}\ and\ \bibinfo {author} {\bibfnamefont {D.}~\bibnamefont {Roy}},\
  }\bibfield  {title} {\emph {\bibinfo {title} {Many-body quantum chaos in
  mixtures of multiple species},\ }}\href {\doibase
  10.1103/PhysRevE.109.L032201} {\bibfield  {journal} {\bibinfo  {journal}
  {Phys. Rev. E}\ }\textbf {\bibinfo {volume} {109}},\ \bibinfo {pages}
  {L032201} (\bibinfo {year} {2024})}\BibitemShut {NoStop}%
\bibitem [{\citenamefont {Ba\~nuls}\ \emph {et~al.}(2011)\citenamefont
  {Ba\~nuls}, \citenamefont {Cirac},\ and\ \citenamefont
  {Hastings}}]{Banuls2011}%
  \BibitemOpen
  \bibfield  {author} {\bibinfo {author} {\bibfnamefont {M.~C.}\ \bibnamefont
  {Ba\~nuls}}, \bibinfo {author} {\bibfnamefont {J.~I.}\ \bibnamefont {Cirac}},
  \ and\ \bibinfo {author} {\bibfnamefont {M.~B.}\ \bibnamefont {Hastings}},\
  }\bibfield  {title} {\emph {\bibinfo {title} {Strong and weak thermalization
  of infinite nonintegrable quantum systems},\ }}\href {\doibase
  10.1103/PhysRevLett.106.050405} {\bibfield  {journal} {\bibinfo  {journal}
  {Phys. Rev. Lett.}\ }\textbf {\bibinfo {volume} {106}},\ \bibinfo {pages}
  {050405} (\bibinfo {year} {2011})}\BibitemShut {NoStop}%
\bibitem [{\citenamefont {Rodriguez-Nieva}\ \emph {et~al.}(2023)\citenamefont
  {Rodriguez-Nieva}, \citenamefont {Jonay},\ and\ \citenamefont
  {Khemani}}]{RodriguezNieva2023}%
  \BibitemOpen
  \bibfield  {author} {\bibinfo {author} {\bibfnamefont {J.~F.}\ \bibnamefont
  {Rodriguez-Nieva}}, \bibinfo {author} {\bibfnamefont {C.}~\bibnamefont
  {Jonay}}, \ and\ \bibinfo {author} {\bibfnamefont {V.}~\bibnamefont
  {Khemani}},\ }\bibfield  {title} {\emph {\bibinfo {title} {Quantifying
  quantum chaos through microcanonical distributions of entanglement},\
  }}\href@noop {} {\bibfield  {journal} {\bibinfo  {journal} {arXiv preprint
  arXiv:2305.11940}\ } (\bibinfo {year} {2023})}\BibitemShut {NoStop}%
\bibitem [{dia()}]{diag}%
  \BibitemOpen
  \href@noop {} {}\bibinfo {note} {{Although in principle one needs to consider
  an operator ${O}$ with its smooth diagonal part (in the energy eigenbasis)
  subtracted, for small $\Delta E$, the contribution of the non-trivial part of
  $O(E)\delta_{mn}$ to the cumulants ${\Delta}_k$ will be suppressed by the
  system size. Thus, in our numerical analysis we use the original observable
  $O$.}}\BibitemShut {Stop}%
\bibitem [{\citenamefont {Heitmann}\ \emph {et~al.}(2020)\citenamefont
  {Heitmann}, \citenamefont {Richter}, \citenamefont {Schubert},\ and\
  \citenamefont {Steinigeweg}}]{Heitmann2020}%
  \BibitemOpen
  \bibfield  {author} {\bibinfo {author} {\bibfnamefont {T.}~\bibnamefont
  {Heitmann}}, \bibinfo {author} {\bibfnamefont {J.}~\bibnamefont {Richter}},
  \bibinfo {author} {\bibfnamefont {D.}~\bibnamefont {Schubert}}, \ and\
  \bibinfo {author} {\bibfnamefont {R.}~\bibnamefont {Steinigeweg}},\
  }\bibfield  {title} {\emph {\bibinfo {title} {Selected applications of
  typicality to real-time dynamics of quantum many-body systems},\ }}\href@noop
  {} {\bibfield  {journal} {\bibinfo  {journal} {Zeitschrift f{\"u}r
  Naturforschung A}\ }\textbf {\bibinfo {volume} {75}},\ \bibinfo {pages} {421}
  (\bibinfo {year} {2020})}\BibitemShut {NoStop}%
\bibitem [{\citenamefont {Jin}\ \emph {et~al.}(2021)\citenamefont {Jin},
  \citenamefont {Willsch}, \citenamefont {Willsch}, \citenamefont {Lagemann},
  \citenamefont {Michielsen},\ and\ \citenamefont {De~Raedt}}]{Jin2021}%
  \BibitemOpen
  \bibfield  {author} {\bibinfo {author} {\bibfnamefont {F.}~\bibnamefont
  {Jin}}, \bibinfo {author} {\bibfnamefont {D.}~\bibnamefont {Willsch}},
  \bibinfo {author} {\bibfnamefont {M.}~\bibnamefont {Willsch}}, \bibinfo
  {author} {\bibfnamefont {H.}~\bibnamefont {Lagemann}}, \bibinfo {author}
  {\bibfnamefont {K.}~\bibnamefont {Michielsen}}, \ and\ \bibinfo {author}
  {\bibfnamefont {H.}~\bibnamefont {De~Raedt}},\ }\bibfield  {title} {\emph
  {\bibinfo {title} {Random state technology},\ }}\href
  {https://doi.org/10.7566/JPSJ.90.012001} {\bibfield  {journal} {\bibinfo
  {journal} {Journal of the Physical Society of Japan}\ }\textbf {\bibinfo
  {volume} {90}},\ \bibinfo {pages} {012001} (\bibinfo {year}
  {2021})}\BibitemShut {NoStop}%
\bibitem [{\citenamefont {Hahn}\ \emph {et~al.}(2023)\citenamefont {Hahn},
  \citenamefont {Luitz},\ and\ \citenamefont {Chalker}}]{Hahn2023}%
  \BibitemOpen
  \bibfield  {author} {\bibinfo {author} {\bibfnamefont {D.}~\bibnamefont
  {Hahn}}, \bibinfo {author} {\bibfnamefont {D.~J.}\ \bibnamefont {Luitz}}, \
  and\ \bibinfo {author} {\bibfnamefont {J.}~\bibnamefont {Chalker}},\
  }\bibfield  {title} {\emph {\bibinfo {title} {The statistical properties of
  eigenstates in chaotic many-body quantum systems},\ }}\href@noop {}
  {\bibfield  {journal} {\bibinfo  {journal} {arXiv preprint arXiv:2309.12982}\
  } (\bibinfo {year} {2023})}\BibitemShut {NoStop}%
\bibitem [{\citenamefont {Belin}\ \emph {et~al.}(2022)\citenamefont {Belin},
  \citenamefont {de~Boer},\ and\ \citenamefont {Liska}}]{Belin:2021ryy}%
  \BibitemOpen
  \bibfield  {author} {\bibinfo {author} {\bibfnamefont {A.}~\bibnamefont
  {Belin}}, \bibinfo {author} {\bibfnamefont {J.}~\bibnamefont {de~Boer}}, \
  and\ \bibinfo {author} {\bibfnamefont {D.}~\bibnamefont {Liska}},\ }\bibfield
   {title} {\emph {\bibinfo {title} {Non-gaussianities in the statistical
  distribution of heavy ope coefficients and wormholes},\ }}\href
  {https://doi.org/10.1007/JHEP06(2022)116} {\bibfield  {journal} {\bibinfo
  {journal} {Journal of High Energy Physics}\ }\textbf {\bibinfo {volume}
  {2022}},\ \bibinfo {pages} {1} (\bibinfo {year} {2022})}\BibitemShut
  {NoStop}%
\bibitem [{\citenamefont {Anous}\ \emph {et~al.}(2022)\citenamefont {Anous},
  \citenamefont {Belin}, \citenamefont {de~Boer},\ and\ \citenamefont
  {Liska}}]{Anous:2021caj}%
  \BibitemOpen
  \bibfield  {author} {\bibinfo {author} {\bibfnamefont {T.}~\bibnamefont
  {Anous}}, \bibinfo {author} {\bibfnamefont {A.}~\bibnamefont {Belin}},
  \bibinfo {author} {\bibfnamefont {J.}~\bibnamefont {de~Boer}}, \ and\
  \bibinfo {author} {\bibfnamefont {D.}~\bibnamefont {Liska}},\ }\bibfield
  {title} {\emph {\bibinfo {title} {Ope statistics from higher-point
  crossing},\ }}\href {https://doi.org/10.1007/JHEP06(2022)102} {\bibfield
  {journal} {\bibinfo  {journal} {Journal of High Energy Physics}\ }\textbf
  {\bibinfo {volume} {2022}},\ \bibinfo {pages} {1} (\bibinfo {year}
  {2022})}\BibitemShut {NoStop}%
\bibitem [{\citenamefont {Jafferis}\ \emph
  {et~al.}(2023{\natexlab{a}})\citenamefont {Jafferis}, \citenamefont
  {Kolchmeyer}, \citenamefont {Mukhametzhanov},\ and\ \citenamefont
  {Sonner}}]{Jafferis:2022uhu}%
  \BibitemOpen
  \bibfield  {author} {\bibinfo {author} {\bibfnamefont {D.~L.}\ \bibnamefont
  {Jafferis}}, \bibinfo {author} {\bibfnamefont {D.~K.}\ \bibnamefont
  {Kolchmeyer}}, \bibinfo {author} {\bibfnamefont {B.}~\bibnamefont
  {Mukhametzhanov}}, \ and\ \bibinfo {author} {\bibfnamefont {J.}~\bibnamefont
  {Sonner}},\ }\bibfield  {title} {\emph {\bibinfo {title} {Matrix models for
  eigenstate thermalization},\ }}\href {\doibase 10.1103/PhysRevX.13.031033}
  {\bibfield  {journal} {\bibinfo  {journal} {Phys. Rev. X}\ }\textbf {\bibinfo
  {volume} {13}},\ \bibinfo {pages} {031033} (\bibinfo {year}
  {2023}{\natexlab{a}})}\BibitemShut {NoStop}%
\bibitem [{\citenamefont {Jafferis}\ \emph
  {et~al.}(2023{\natexlab{b}})\citenamefont {Jafferis}, \citenamefont
  {Kolchmeyer}, \citenamefont {Mukhametzhanov},\ and\ \citenamefont
  {Sonner}}]{Jafferis:2022wez}%
  \BibitemOpen
  \bibfield  {author} {\bibinfo {author} {\bibfnamefont {D.~L.}\ \bibnamefont
  {Jafferis}}, \bibinfo {author} {\bibfnamefont {D.~K.}\ \bibnamefont
  {Kolchmeyer}}, \bibinfo {author} {\bibfnamefont {B.}~\bibnamefont
  {Mukhametzhanov}}, \ and\ \bibinfo {author} {\bibfnamefont {J.}~\bibnamefont
  {Sonner}},\ }\bibfield  {title} {\emph {\bibinfo {title} {Jackiw-teitelboim
  gravity with matter, generalized eigenstate thermalization hypothesis, and
  random matrices},\ }}\href {\doibase 10.1103/PhysRevD.108.066015} {\bibfield
  {journal} {\bibinfo  {journal} {Phys. Rev. D}\ }\textbf {\bibinfo {volume}
  {108}},\ \bibinfo {pages} {066015} (\bibinfo {year}
  {2023}{\natexlab{b}})}\BibitemShut {NoStop}%
\bibitem [{\citenamefont {Belin}\ \emph {et~al.}(2023)\citenamefont {Belin},
  \citenamefont {de~Boer}, \citenamefont {Jafferis}, \citenamefont {Nayak},\
  and\ \citenamefont {Sonner}}]{Belin:2023efa}%
  \BibitemOpen
  \bibfield  {author} {\bibinfo {author} {\bibfnamefont {A.}~\bibnamefont
  {Belin}}, \bibinfo {author} {\bibfnamefont {J.}~\bibnamefont {de~Boer}},
  \bibinfo {author} {\bibfnamefont {D.~L.}\ \bibnamefont {Jafferis}}, \bibinfo
  {author} {\bibfnamefont {P.}~\bibnamefont {Nayak}}, \ and\ \bibinfo {author}
  {\bibfnamefont {J.}~\bibnamefont {Sonner}},\ }\bibfield  {title} {\emph
  {\bibinfo {title} {Approximate cfts and random tensor models},\ }}\href@noop
  {} {\bibfield  {journal} {\bibinfo  {journal} {arXiv preprint
  arXiv:2308.03829}\ } (\bibinfo {year} {2023})}\BibitemShut {NoStop}%
\bibitem [{\citenamefont {Forrester}(2006)}]{forrester2006quantum}%
  \BibitemOpen
  \bibfield  {author} {\bibinfo {author} {\bibfnamefont {P.~J.}\ \bibnamefont
  {Forrester}},\ }\bibfield  {title} {\emph {\bibinfo {title} {Quantum
  conductance problems and the jacobi ensemble},\ }}\href {\doibase
  10.1088/0305-4470/39/22/004} {\bibfield  {journal} {\bibinfo  {journal}
  {Journal of Physics A: Mathematical and General}\ }\textbf {\bibinfo {volume}
  {39}},\ \bibinfo {pages} {6861} (\bibinfo {year} {2006})}\BibitemShut
  {NoStop}%
\bibitem [{\citenamefont {Savin}\ \emph {et~al.}(2008)\citenamefont {Savin},
  \citenamefont {Sommers},\ and\ \citenamefont
  {Wieczorek}}]{savin2008nonlinear}%
  \BibitemOpen
  \bibfield  {author} {\bibinfo {author} {\bibfnamefont {D.~V.}\ \bibnamefont
  {Savin}}, \bibinfo {author} {\bibfnamefont {H.-J.}\ \bibnamefont {Sommers}},
  \ and\ \bibinfo {author} {\bibfnamefont {W.}~\bibnamefont {Wieczorek}},\
  }\bibfield  {title} {\emph {\bibinfo {title} {Nonlinear statistics of quantum
  transport in chaotic cavities},\ }}\href {\doibase
  10.1103/PhysRevB.77.125332} {\bibfield  {journal} {\bibinfo  {journal} {Phys.
  Rev. B}\ }\textbf {\bibinfo {volume} {77}},\ \bibinfo {pages} {125332}
  (\bibinfo {year} {2008})}\BibitemShut {NoStop}%
\bibitem [{\citenamefont {Vidal}\ and\ \citenamefont
  {Kanzieper}(2012)}]{vidal2012statistics}%
  \BibitemOpen
  \bibfield  {author} {\bibinfo {author} {\bibfnamefont {P.}~\bibnamefont
  {Vidal}}\ and\ \bibinfo {author} {\bibfnamefont {E.}~\bibnamefont
  {Kanzieper}},\ }\bibfield  {title} {\emph {\bibinfo {title} {Statistics of
  reflection eigenvalues in chaotic cavities with nonideal leads},\ }}\href
  {\doibase 10.1103/PhysRevLett.108.206806} {\bibfield  {journal} {\bibinfo
  {journal} {Phys. Rev. Lett.}\ }\textbf {\bibinfo {volume} {108}},\ \bibinfo
  {pages} {206806} (\bibinfo {year} {2012})}\BibitemShut {NoStop}%
\bibitem [{\citenamefont {Puchala}\ and\ \citenamefont
  {Miszczak}(2017)}]{Pucha_a_2017}%
  \BibitemOpen
  \bibfield  {author} {\bibinfo {author} {\bibfnamefont {Z.}~\bibnamefont
  {Puchala}}\ and\ \bibinfo {author} {\bibfnamefont {J.}~\bibnamefont
  {Miszczak}},\ }\bibfield  {title} {\emph {\bibinfo {title} {Symbolic
  integration with respect to the haar measure on the unitary groups},\ }}\href
  {\doibase 10.1515/bpasts-2017-0003} {\bibfield  {journal} {\bibinfo
  {journal} {Bulletin of the Polish Academy of Sciences Technical Sciences}\
  }\textbf {\bibinfo {volume} {65}},\ \bibinfo {pages} {21} (\bibinfo {year}
  {2017})}\BibitemShut {NoStop}%
\bibitem [{\citenamefont {Wei\ss{}e}\ \emph {et~al.}(2006)\citenamefont
  {Wei\ss{}e}, \citenamefont {Wellein}, \citenamefont {Alvermann},\ and\
  \citenamefont {Fehske}}]{Fehske06}%
  \BibitemOpen
  \bibfield  {author} {\bibinfo {author} {\bibfnamefont {A.}~\bibnamefont
  {Wei\ss{}e}}, \bibinfo {author} {\bibfnamefont {G.}~\bibnamefont {Wellein}},
  \bibinfo {author} {\bibfnamefont {A.}~\bibnamefont {Alvermann}}, \ and\
  \bibinfo {author} {\bibfnamefont {H.}~\bibnamefont {Fehske}},\ }\bibfield
  {title} {\emph {\bibinfo {title} {The kernel polynomial method},\ }}\href
  {\doibase 10.1103/RevModPhys.78.275} {\bibfield  {journal} {\bibinfo
  {journal} {Rev. Mod. Phys.}\ }\textbf {\bibinfo {volume} {78}},\ \bibinfo
  {pages} {275} (\bibinfo {year} {2006})}\BibitemShut {NoStop}%
\bibitem [{\citenamefont {Osborn}\ and\ \citenamefont
  {Verbaarschot}(1998)}]{Osborn:1998nf}%
  \BibitemOpen
  \bibfield  {author} {\bibinfo {author} {\bibfnamefont {J.~C.}\ \bibnamefont
  {Osborn}}\ and\ \bibinfo {author} {\bibfnamefont {J.~J.~M.}\ \bibnamefont
  {Verbaarschot}},\ }\bibfield  {title} {\emph {\bibinfo {title} {Thouless
  energy and correlations of qcd dirac eigenvalues},\ }}\href {\doibase
  10.1103/PhysRevLett.81.268} {\bibfield  {journal} {\bibinfo  {journal} {Phys.
  Rev. Lett.}\ }\textbf {\bibinfo {volume} {81}},\ \bibinfo {pages} {268}
  (\bibinfo {year} {1998})}\BibitemShut {NoStop}%
\bibitem [{\citenamefont {Srednicki}(1999)}]{Srednicki1999}%
  \BibitemOpen
  \bibfield  {author} {\bibinfo {author} {\bibfnamefont {M.}~\bibnamefont
  {Srednicki}},\ }\bibfield  {title} {\emph {\bibinfo {title} {The approach to
  thermal equilibrium in quantized chaotic systems},\ }}\href {\doibase
  10.1088/0305-4470/32/7/007} {\bibfield  {journal} {\bibinfo  {journal}
  {Journal of Physics A: Mathematical and General}\ }\textbf {\bibinfo {volume}
  {32}},\ \bibinfo {pages} {1163} (\bibinfo {year} {1999})}\BibitemShut
  {NoStop}%
\end{thebibliography}%

\clearpage
\newpage
\setcounter{figure}{0}
\setcounter{equation}{0}
\renewcommand{\thefigure}{S\arabic{figure}}
\renewcommand{\theequation}{S\arabic{equation}}

\section*{Supplemental material}

	\subsection*{Random unitary projector}
	Our starting point is a large $N\times N$ matrix $A$ (with $N\gg1$), projected onto an $M$-dimensional subspace with a Haar-random unitary projector. We can write that explicitly as 
	\begin{equation}
	\tilde{A}=P\, U\, A\, U^\dagger P^T, 
	\end{equation}
	where $U$ is an $N\times N$ Haar-random unitary and $P$ is a fixed projector of rank $M$, an $M\times N$ matrix satisfying $P P^T={\mathcal I}_M$. 
	{Although not necessary for what follows, we can introduce an $M\times M$ Haar-random unitary $V$ with the substitution $P\rightarrow V P$, 
		such that resulting $M\times M$ matrix $\tilde{A}$ will exhibit unitary symmetry, and the spectrum will be its only invariant. The combination $VPU$ is a random unitary projector.}
	
	For simplicity we can choose $P$ to be the projector on first $M$ coordinates
	\begin{equation}
	\tilde{A}_{ij}=U_i^k A_{kl} (U^*)_{j}^l\ ,\quad 1\leq i,j\leq M, \quad 1\leq k,l\leq N\ .
	\end{equation}
	At this point it is convenient to use block representation of the Haar-random unitary (orthogonal) matrix $U$, which is well-known in physics literature modeling scattering matrix of 1D systems \cite{forrester2006quantum, savin2008nonlinear, vidal2012statistics}, 
	\begin{eqnarray}
	\label{decomposition}
	U=\left(
	\begin{array}{c|c}
	r_{M\times M} & t_{M\times (N-M)} \\
	\hline
	t'_{(N-M) \times  M} & r'_{(N-M)\times (N-M)}
	\end{array}
	\right)\ ,
	\end{eqnarray}
	Here, 
	\begin{eqnarray}
	r_{M\times M}=u_1\, {\rm diag}(1-T_i)\, u_2^\dagger\ ,\\
	t_{M\times (N-M)}= i u_1  \Lambda v_2^\dagger,\qquad \Lambda_{ij}=\delta_{ij}T_i^{1/2}\ ,
	\end{eqnarray}
	where $u_1,u_2$ are two Haar-random $M\times M$ unitary matrices and $v_1,v_2$ are two random $(N-M)\times (N-M)$ unitary matrices. Moreover, $T_i=0$ for $i>M$, while for $i\leq M$, $0\leq T_i\leq 1$ are random variables with a joint probability distribution,
	\begin{equation}
	\label{measure}
	dP={1\over S}\prod_{i=1}^M dT_i\, T_i^{\alpha-1} \prod_{i<j}|T_i-T_j|^\beta\ ,     
	\end{equation}
	with $\alpha=\beta(N-2M+1)/2$, and $\beta=1,2,4$ as usual marks GOE, GUE, or GSE ensemble (meaning $U$ and other matrices above may have been Haar-orthogonal or Haar-symplectic). Above we have also assumed that $M\leq N/2$. Factor $S$ in \eqref{measure} is added for normalization, it can be evaluated explicitly using Selberg integral,
	\begin{eqnarray}
	\nonumber
	S(\tilde{\alpha},\tilde{\beta},\tilde{\gamma})&=&\prod_{i=1}^M dT_i\, T_i^{\tilde\alpha-1} (1-T_i)^{\tilde{\beta}-1} \prod_{i<j}|T_i-T_j|^{2\tilde\gamma}\\
	&&=\prod_{j=0}^{M-1} {\Gamma(\tilde{\alpha}+j\tilde{\gamma})\Gamma(\tilde{\beta}+j\tilde{\gamma})\Gamma(1+(j+1)\tilde{\gamma})\over \Gamma(\tilde{\alpha}+\tilde{\beta}+(M+j-1)\tilde{\gamma})\Gamma(1+\tilde{\gamma})}\ ,
	\nonumber
	\end{eqnarray}
	with the parameters $\tilde\alpha=\alpha,\, \tilde\beta=1,\, \tilde{\gamma}=\beta/2$. The expectation value of the product $T_1 T_2\dots T_k$ with the measure \eqref{measure} is given by Aomoto's integral formula,
	\begin{eqnarray}
	&&\prod_{i=1}^M dT_i\, T_i^{\tilde\alpha-1} \left(\prod_{i=1}^k T_i\right) (1-T_i)^{\tilde{\beta}-1} \prod_{i<j}|T_i-T_j|^{2\tilde\gamma} \nonumber\\ && \qquad =S\prod_{j=1}^k {\tilde{\alpha}+(M-j)\tilde{\gamma}\over \tilde{\alpha}+\tilde{\beta}+(2M-j-1)\gamma}\ .
	\end{eqnarray}
	After changing the variables $T_i\rightarrow 1-T_i$ we find, 
	\begin{equation}
	\label{expvalue}
	\langle \prod_{i=1}^k (1-T_i)\rangle={\tilde{M}!\, (\tilde{N}-k)!\over \tilde{N}!\, (\tilde{M}-k)!}\ ,
	\end{equation}
	where $\tilde{M}=M+2/\beta-1$, and $\tilde{N}=N+2/\beta-1$. Since we are working in the limit of large $N\geq M\gg 1$ the difference between $\tilde{M},\tilde{N}$ and $M,N$ and hence the difference between different Gaussian ensembles (unitary, orthogonal or symplectic) can be neglected. Our goal is to evaluate moments of $\tilde{A}$
	\begin{equation}\label{M2k}
	{\mathcal M}_k={\langle {\rm Tr}\tilde{A}^k\rangle\over M}\ ,
	\end{equation}
	starting from the known moments of $A$, ${\mathcal A}_k={\rm Tr}(A^k)/N$. As a starting point we evaluate,  
	\begin{equation}
	{\rm det}(z-\tilde{A})=z^M e^{{\rm Tr}\ln(1-\tilde{A}/z)}\ .
	\end{equation}
	Without loss of generality we can assume matrix $A$ is diagonal, with the eigenvalues $a_i$. Furthermore, as a technical assumption, we consider the case when all $a_i=0$ for $i>M$. It is easy to see then that each term in $1/z$ expansion of
	\begin{eqnarray}
	&& e^{{\rm Tr}\ln(1-\tilde{A}/z)}=\\ 
	&& \quad 1-{\sum_i (1-T_i)a_i\over z}+{\sum_{i\neq j} (1-T_i)(1-T_j)a_ia_j/2\over  z^2}+\dots \nonumber
	\end{eqnarray}
	only involve products of $1-T_i$ with distinct indexes. Using \eqref{expvalue} we readily find,
	\begin{eqnarray}
	\langle {\rm det}(z-\tilde{A})\rangle=z^M \sum_{k=0}^\infty  {a_k\over z^k}  {\tilde{M}!\, (\tilde{N}-k)!\over \tilde{N}!\, (\tilde{M}-k)!}\ , 
	\end{eqnarray}
	where $a_k$ are defined via,
	\begin{eqnarray}
	&&{\rm exp}\left(-\sum_{k=1}^\infty {{\rm Tr}(A^k)\over k\, z^k} \right)\equiv 1+ \sum_{k=1}^\infty {a_k\over z^k},\\
	&& a_1=-{\rm Tr}A,\quad a_2={({\rm Tr}\,A)^2-{\rm Tr}\, A^2\over 2}\ , \quad \dots \nonumber
	\end{eqnarray}
	
	Eventually, we are interested in calculating resolvent 
	\begin{eqnarray}
	{1\over M}\langle {\rm Tr}{1\over z-\tilde{A}}\rangle={1\over z}+\sum_{k=1}^\infty {{\mathcal M_k}\over z^{k+1}}\ , 
	\end{eqnarray}
	which is given by $z$-derivative of $\langle \ln {\rm det}(z-\tilde{A})\rangle$. In the large-$M$ limit we can instead evaluate $z$-derivative of $\ln\,\langle{ \rm det}(z-\tilde{A}) \rangle $ yielding,
	\begin{eqnarray}
	\sum_{k=1}^\infty {{\mathcal M_k}\over z^{k+1}} ={1\over M}{d\over dz}\ln \left(
	1+ \sum_{k=1}^\infty {a_k\over z^k} {{M}!\, ({N}-k)!\over {N}!\, ({M}-k)!}
	\right)\ . \nonumber
	\end{eqnarray}
	We have removed tilde from $M$ and $N$ because we are working in the large-$M,N$ limit. 
	After introducing $\alpha=M/N$ in the large-$M,N$ limit we find
	\begin{eqnarray}
	\nonumber
	&&   {\mathcal M}_1={\mathcal A}_1,\\
	\nonumber
	&&  {\mathcal M}_2=\alpha {\mathcal A}_2+(1-\alpha){\mathcal A}_1^2,\\
	\nonumber
	&& {\mathcal M}_3=\alpha^2 {\mathcal A}_3+3\alpha(1-\alpha) {\mathcal A}_1{\mathcal A}_2+(2\alpha^2-3\alpha+1){\mathcal A}_1^3,\\
	\nonumber 
	&& {\mathcal M}_4=\alpha^3 {\mathcal A}_4+4\alpha^2(1-\alpha)  {\mathcal A}_1{\mathcal A}_3+2\alpha^2(1-\alpha)  {\mathcal A}_2^2+ \\
	\nonumber && 2\alpha(5\alpha^2-8\alpha+3){\mathcal A}_1^2 {\mathcal A}_2+ (-5\alpha^3+10\alpha^2-6\alpha+1){\mathcal A}_1^4,\\
	&& \dots \label{eq-M1234},
	\end{eqnarray}
	c.f.~Eqs.\ (7) and (8) in the main text. Here we have dropped the assumption that $A$ has at most $M$ non-zero eigenvalues. Correctness of final result can be checked for low moments directly by writing it in components, e.g.,
	\begin{eqnarray}
	M\, {\mathcal M}_2=\sum_{i,j=1}^M   \sum_{k,l,r,t=1}^N  \langle U_{ik}A_{kl}U^*_{lj}U_{jr}A_{rt}U^*_{ti} \rangle\ ,
	\end{eqnarray}
	and using multi-point correlations of individual matrix elements $U_{ij}$ \cite{Pucha_a_2017}.
	At this point we introduce free cumulants $\Delta_k$ 
	\begin{align}
	\Delta_1 &= {\cal M}_1 \nonumber \\
	\Delta_{2}&={\cal M}_{2}-{\cal M}_{1}^{2} \nonumber \\
	\Delta_{3}&={\cal M}_{3}-3{\cal M}_{2}{\cal M}_{1}+2{\cal M}_{1}^{3} \nonumber \\
	\Delta_{4}& ={\cal M}_{4}-4{\cal M}_{3}{\cal M}_{1}-2{\cal M}_{2}^{2}+10{\cal M}_{2}{\cal M}_{1}^{2}-5{\cal M}_{1}^{4} \nonumber \\
	&\vdots \label{eq-Delta1234}
	\end{align}
	After a straightforward calculation  Eq. \eqref{eq-M1234} yields
	\begin{equation}\label{eq-fc-result0}
	\Delta_{k}^{(M)}=\alpha^{k-1}\Delta_{k}^{(N)},
	\end{equation}
	where $\Delta_{k}^{(M)}$ and $\Delta_{k}^{(N)}$ indicate free cumulants defined in terms of  moments ${\cal M}$ and $\cal A$, respectively. 
	Eq. \eqref{eq-fc-result0} implies a universal behavior of free cumulants under unitary symmetry.

	\begin{figure}[tb]
		\centering
		\includegraphics[width = 1.0\linewidth]{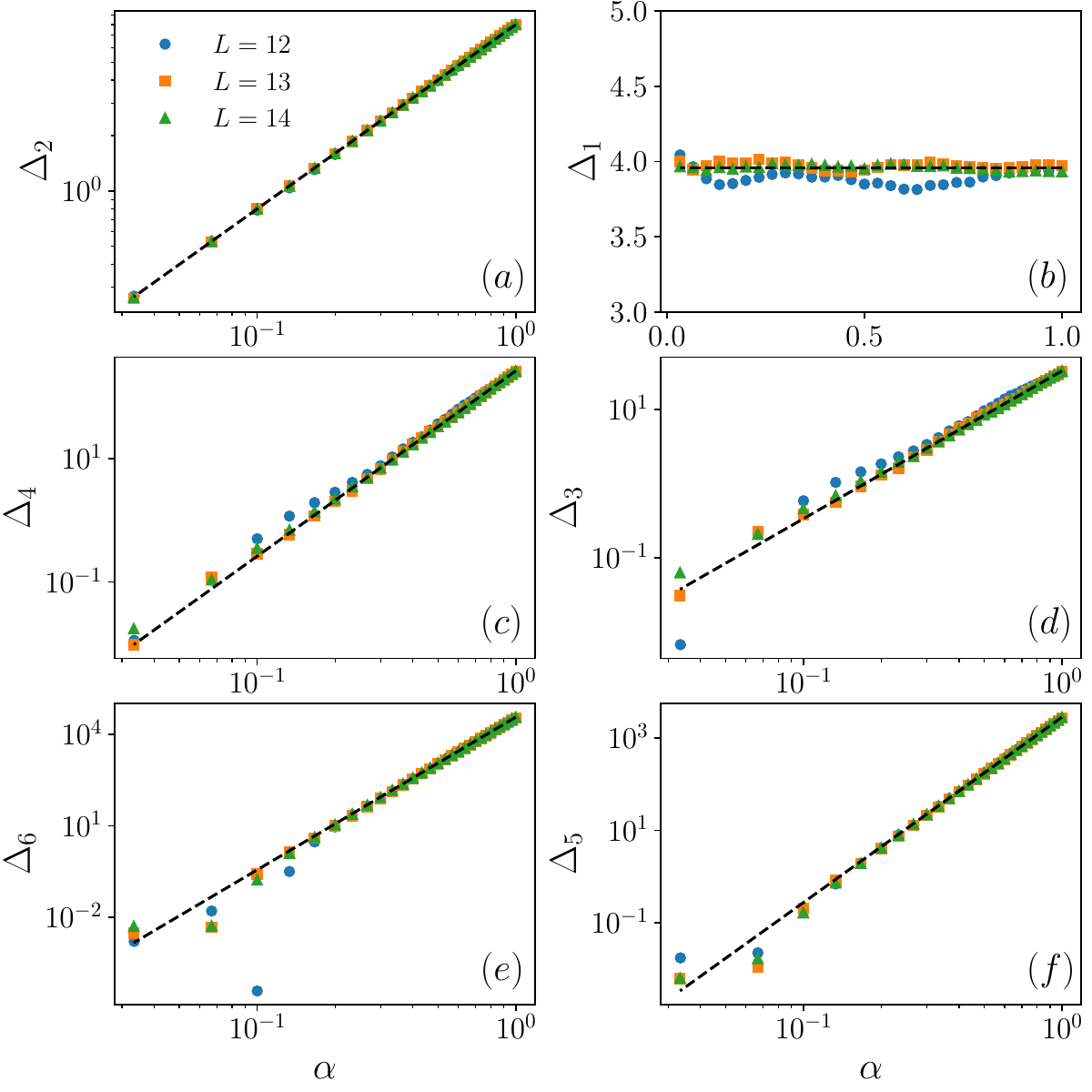}
		\caption{Numerical demonstration of the analytic results for a random-matrix toy model. Free cumulants $\Delta_{k}$ versus $\alpha = \frac{d}{D}$ for an operator ${\cal O}$ whose spectrum follows a $4$th order $\chi^2$ distribution and a Hamiltonian drawn from GOE, leading to the Haar-random orthogonal $U$. Results are shown for Hilbert-space dimension $D=2^L$, where $L=12,13,14$.   
			{As a guide to the eye,  the dashed line indicates the theoretically predicted slope $\ln\Delta_{k}=(k-1)\ln d+{\rm const}$.}
		}\label{Fig-GOE}
	\end{figure}
	
	We provide a numerical check of our analytical result, Eq.~\eqref{eq-fc-result0}. To that end we consider a diagonal random matrix ${\cal O}^*$ from Eq. \ (5) in the main text, with the matrix elements drawn from the $4$th order $\chi^2$ distribution, and the transformation $U$ from a Haar-random orthogonal ensemble (one can think of a Hamiltonian drawn from the  GOE). We evaluate free cumulants $\Delta_k$ numerically and show them as a function  of the  ratio $\alpha=\frac{d}{D}$ (ratio of the energy window's dimension and the whole Hilbert space size) in Fig.~\ref{Fig-GOE}. The scaling $\Delta_k \propto \alpha^{k-1}$ can be clearly observed, except for the fluctuations at the very small window sizes, which are due to insufficient statistics for very small $d$. The results are in agreement with  \eqref{eq-fc-result0}.

	{If we now assume that microcanonically projected operator ${\cal O}_{\Delta E}$ from the main text, 
		for $\Delta E\leq \Delta E_U$ exhibits (emergent) unitary symmetry, and assuming  $\Delta E_U$ is small enough such that density of states is approximately constant  $d(\Delta E)\propto \Delta E$, we find  
		\begin{equation}\label{eq-Deltak-DE}
		\Delta_{k}(\Delta E)=\lambda_{k}\,\Delta E^{k-1}, 
		\end{equation}
		where $\lambda_k=\Delta_k^U /(\Delta E_U)^{k-1}$, $\Delta_k^U$ are $k$-th cumulants of ${\cal O}_U\equiv {\cal O}_{\Delta E_U}$.}

	\subsection*{Relation to general ETH}
	We consider the same theoretical model as above, a unitary-invariant operator 
	\begin{equation}
	{\cal O}_U=U {\cal O}^\ast U^\dagger\ ,
	\end{equation}
	where $U$ is a Haar-random unitary operator of size $d_U\equiv d(\Delta E_U)$.  We denote  the matrix elements of ${\cal O}_U$ simply by ${\cal O}_{ij}$ (subindex $U$ is dropped), and 
	study correlation functions between matrix elements,
	\begin{equation}
	F_{k}(i_{1},\cdots,i_{k}):=\overline{{\mathcal O}_{i_{1}i_{2}}\cdots {\mathcal O}_{i_{k}i_{1}}}\ , 
	\end{equation}
	where averaging is over the unitary group.
	One can readily see that the result is independent of $i_1,\dots, i_k$
	\begin{eqnarray}
	F_k=d_U^{k-1}\Delta_k^U\ ,
	\end{eqnarray}
	where $\Delta_k^U$ is the $k$-th free cumulant of ${\mathcal O}_U$.
	To make a connection with general ETH, we consider an operator ${\mathcal  O}_{mn}$ written in the eigenbasis of a chaotic Hamiltonian $H$. Then the correlation function, 
	\begin{eqnarray}
	\overline{{\mathcal O}_{i_{1}i_{2}}\cdots {\mathcal O}_{i_{k}i_{1}}}=e^{-(k-1)S(E_+)}f_{k}(\vec{\omega})\ , 
	\end{eqnarray}
	where average is understood to be over the values of $i_\alpha$ within some narrow intervals centered around $E_\alpha$, will be a smooth function of its arguments,
	\begin{eqnarray}
	E_+&=&(E_1+\dots + E_k)/k\ ,\\ \vec{\omega}&=&(E_2-E_1,\dots,E_k-E_{k-1})\ ,
	\end{eqnarray}
	This was recently proposed and verified numerically in \cite{PhysRevLett.129.170603, Pappalardi:2023nsj}.
	
	Comparing this with the prediction of unitary invariant model above, we readily conclude that emergent unitary symmetry at scales smaller than $\Delta E_U$ is equivalent to condition that cumulants $f_k(\vec{\omega})$ become constant at small frequencies, 
	when all components of $\vec{\omega}$ are not exceeding $\Delta E_U$,
	\begin{eqnarray}
	f_k(\vec{\omega})= (e^{S(E_+)}/d_U)^{k-1} \Delta^U_k\ . 
	\end{eqnarray}
	Taking into account Eq.\ (9) in the main text, this expression will not change if instead of $\Delta E_U$ one considers any other smaller window $\Delta E<\Delta E_U$. 
	
	For example for $k=2$ we find,
	\begin{eqnarray}
	f^2(\bar E,\omega)=(e^{S(\bar{E})}/d) \Delta_2\ , \quad {\rm for}\, \, |\omega|\leq \Delta E_U\ ,
	\end{eqnarray}
	in full agreement with the ETH ansatz~\cite{Srednicki1999},
	\begin{eqnarray}
	\label{eq::ETH}
	{\mathcal O}_{mn} &=& {\mathcal O}(\bar{E})\delta_{mn} + 
	e^{-\frac{S(\bar{E})}{2}}f(\bar{E},\omega)r_{mn}\  ,\\
	\omega &=& E_m-E_n\ ,\quad \bar{E} = (E_m + E_n)/2\ .
	\end{eqnarray}
	We emphasize, constant value of $f^2(\omega)$ is a necessary but not sufficient condition for unitary symmetry. Full unitary symmetry requires all higher cumulant functions $f_k$ to be constant at small frequencies, when all components of $\vec{\omega}$ are smaller than $\Delta E_U$. This condition was already recognized in \cite{PhysRevLett.129.170603,Pappalardi:2023nsj} as the condition for the ``rotation-invariant ETH'' of \cite{PhysRevLett.123.260601}, which, up to certain subtleties, is mathematically equivalent to the model we considered. The crucial difference between our work and \cite{PhysRevLett.129.170603,Pappalardi:2023nsj,PhysRevLett.123.260601} is in the choice of observables and the numerical method, which allows us to probe the scale when this behavior emege. Evaluating cumulant functions $f_k$ requires exact diagonalization, which limits analysis to modest system sizes. By formulating the implications of unitary symmetry in terms of cumulants $\Delta_k$ we devised the readily testable predictions (Eq.\ (9) in the main text), which can be probed numerically at larger system sizes with help of methods of typicality.  
	%
	
	

	\subsection*{Definitions and the relation between $\Delta E_{Th}, \Delta E_T, \Delta E_U, \Delta E_{GUE}$}
	We start with Thouless energy $\Delta E_{Th}$, which marks the onset of Random Matrix Theory behavior of energy levels. To define $\Delta E_{Th}$ one considers {\it level rigidity}, variance in the number $N$ of energy levels within a narrow microcanonical window of width $\Delta E$, centered around some $E$,
	\begin{equation}
	(\langle N^2\rangle -\langle N\rangle^2)(\Delta E)\ .
	\end{equation}
	Averaging here is understood in terms of choosing different windows, with different but similar central energy $E$, with the same energy density.
	
	For a quantum chaotic system the Bohigas-Giannoni-Schmit conjecture \cite{Bohigas} predicts that for sufficiently small $\Delta E$ the answer of  level rigidity would be the same as given by a Gaussian RMT. Say, in case of GUE, the variance would be proportional to $\ln(\Delta E)$. Thouless energy is defined as the smallest $\Delta E$, for which variance will start deviating from the RMT value. To quantify that, one normally introduces a small parameter $\epsilon$, such that 
	\begin{eqnarray}
	\label{Thdef}
	\left|{\langle N^2\rangle_{RMT}  \over \langle N^2\rangle}-1\right|_{\Delta E_{Th}}=\epsilon\ .
	\end{eqnarray}
	A similar definition can be given in time domain in terms of the spectral form factor. 
	
	For spatially-extended systems with local interaction Thouless energy will be decreasing (non increasing) with system size, while the  density of states $e^S$ will be a function of energy density $E/V$, where $V$ is the volume. Thus, for sufficiently large systems the definition of $\Delta E_{Th}$ would not require spectrum unfolding, which is a standard numerical procedure to define $\Delta E_{Th}$ for finite $V$ \cite{Osborn:1998nf}.
	
	The definition above is ambiguous, it depends on an arbitrary small parameter $\epsilon$. The idea is to keep $\epsilon$ fixed while taking system size $V$ to be very large. In this case $\Delta E_{Th}$ defines a {\it scale}, together with its system-size dependence. It is expected that the inverse Thouless energy $1/\Delta E_{Th}$ will be the timescale of the slowest transport mode \cite{altshuler1986repulsion,PhysRevLett.123.210603,PhysRevX.12.021009,PhysRevB.105.104509,winer2023emergent,winer2023reappearance}. For a chaotic 1D system of the kind discussed in the main text, we expect the slowest transport mode to be diffusive, yielding $\Delta E_{Th}\sim L^{-2}$, where $L$ is the system length. 
	
	Next we discuss inverse thermalization time $E_T$. Unlike $\Delta E_{Th}$, which was a property of Hamiltonian, inverse thermalization time, as well as scales $\Delta E_U$ and $\Delta E_{GUE}$ introduced below, are operator-specific. We define $E_T$ as an inverse thermalization time, where the latter is time $T$ which marks the saturation of the (integral of the) 2pt-function of $\mathcal O$. In the energy domain this is the same as the size of the plateau region of $f^2$ \eqref{eq::ETH} at small $\omega$. In practice to define $E_T$ we use free cumulant $\Delta_2(\Delta E)$. As discussed above, constant $f^2(\omega)$ is equivalent to linearly growing $\Delta_2(\Delta E)$, we thus define $\Delta E_T$ as the minimal value of $\Delta E$ for which $\Delta_2$ deviates from linear growth 
	\begin{eqnarray}
	\label{eq:ETdef}
	\left|\frac{\Delta_{2}}{a\ \Delta E}-1\right|_{\Delta E_{T}}=\epsilon\ , 
	\end{eqnarray}
	and $a\, \Delta E$ is the best fit for $\Delta_2(\Delta E)$ at small $\Delta E$. 
	
	As in the case of Thouless energy we had to introduce an arbitrary small parameter $\epsilon$ [not related to $\epsilon$ in \eqref{Thdef}], and assume that system size $V$ is very large, such that one can neglect nonlinearities of the energy spectrum density. This definition provides a rigorous way, at least in principle, to define $\Delta E_T$ scaling with $V$. As we mentioned before, it is expected that  $\Delta E_{Th}$ will match $\Delta E_T$ for the slowest operator (i.e. smallest $\Delta E_T$ among all local operators) in the sense of system-size dependence, $E_{Th}\sim E_T$. Thus, for the spin chain discussed in the text, which we expect to be diffusive, for typical local operators we expect $E_T \sim L^{-2}$.  For the density-wave operators (Eq. (11) in the main text)  we expect
	\begin{eqnarray}
	\Delta E_{T}\sim {q^2\over L^2}\ ,
	\end{eqnarray}
	where $q$ controls the wave length. 
	Our numerical data is too sensitive to finite size corrections to verify this scaling numerically, as we now explain.
	
\begin{figure}[tb]
		\includegraphics[width=0.8\linewidth]{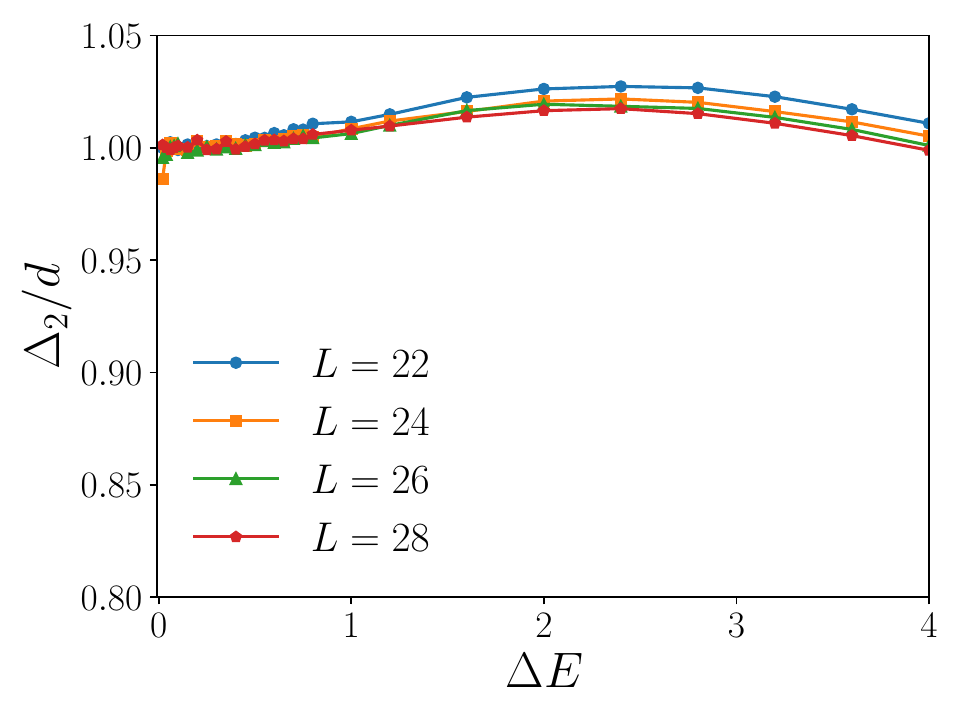}
		\caption{$\Delta_2/d$ versus $\Delta E$ for the operator ${\cal A}_{q=\frac{L}{2}}$. This operator, in the large system size limit, should have $L$-independent thermalization time. This behavior is difficult to observe from the available data for $\Delta_2$. The data confirms, that for small $\Delta E$ the ratio $\Delta_2/d$ is approximately constant (the plateau of $f^2(\omega)$). Now, if we define $\Delta E_T$ as the point where $\Delta_2/d$ deviates, say by $1\%$, from the constant value, the approximate numerical value will be $\Delta E_T\approx 0.3$. But since the curve of $(\Delta_2/d)$ changes its overall shape for growing $L$, it is conceivable that for large L the ``plateau'' region will extend to much large values of order $\Delta E_T\approx 1-1.5$. }\label{Fig-GW}
	\end{figure}

	We note, that the definition of $E_T$ in terms of $\Delta_2$, while conceptually straightforward,  is  prone 
	to significant finite size effects. This is because the shape of $f^2(\omega)$ changes significantly for small $L$, and therefore the plateau region has to be either defined with a very large ``tolerance'' $\epsilon$ or $\Delta E_T$ can change drastically as $L$ increases. We illustrate the problem in Fig.~\ref{Fig-GW}, where $\Delta_2/d$ (essentially $\int_0^{\Delta E}f^2(\omega) d\omega/\Delta E$) is shown as a function of $\Delta E$ for different $L$.
	Finite size effects in the definition of $\Delta E_T$ is one of the main factors complicating finite scaling analysis of  $\Delta E_U/\Delta E_T$.

	Finally, we define $\Delta E_U$ -- the central concept introduced in this paper -- as the scale marking the onset of emergent unitary symmetry. Namely we require that for $\Delta E\leq \Delta E_U$  free cumulants scale as outlined in \eqref{eq-fc-result0}. Again, we can introduce arbitrary small $\epsilon$ and define a energy scale $\Delta E^{(k)}_U$ as the point of deviation from the unitary invariance-imposed behavior for $\Delta_k$, cf.~\eqref{eq:ETdef},
	\begin{eqnarray}
	\label{eq:EUdef}
	\left|\frac{\Delta_{k}}{a_{k}(\Delta E)^{k-1}}-1\right|_{\Delta E^{(k)}_{U}}=\epsilon\ ,  
	\end{eqnarray}
	where $a_k (\Delta E)^{k-1}$ is the best fit for $\Delta_k$ for small $\Delta E$. As above, as the system size grows, we can neglect nonlinearities of the energy spectrum density distinguishing $(\Delta E)^{k-1}$ from $d^{k-1}$.
	We notice that in all cases considered, $\Delta E_U^{(l)} \le \Delta E_U^{(k)}$ for $l<k$. Therefore 
	we can equivalently say that  for $\Delta E\leq \Delta E_U^{(k)}$, all cumulants $\Delta_l$ with $l\leq k$ deviate from Eq. \eqref{eq-fc-result0} by not more than some $\epsilon$.

	
	%
	%
	
	
	With this definition we readily find $\Delta E_U^{(2)}=\Delta E_T$. In our numerical analysis we defined $\Delta E_U=\Delta E_U^{(4)}$ and saw that in most cases $\Delta E_U^{(6)} \approx \Delta E_U^{(4)}\ll \Delta E_U^{(2)}=\Delta E_T$.
	An interesting question is to understand the scaling of $\Delta E_U^{(k)}$ for $k\gg 1$. To this end, we introduce one more scale $\Delta E_{GUE}$ which marks the onset of Gaussian RMT universality. Gaussian RMT is characterized by vanishing higher free cumulants, $\Delta_k=0$, except for $k=2$. It is natural thus to define $\Delta E_{GUE}^{(k)}$, similarly to \eqref{eq:EUdef}, as a maximal value of $\Delta E$ within which all higher moments, properly normalized, plunge below some $\epsilon$,
	\begin{eqnarray}
	\label{eq:ERMTdef}
	\left|{\Delta_k \over (\Delta_2)^{k/2}}\right|_{\Delta E_{GUE}}=\epsilon\ .
	\end{eqnarray}
	So far the operator exhibits emergent unitary symmetry, \eqref{eq-fc-result0} predicts 
	\begin{eqnarray}
	\frac{\Delta_{k}}{(\Delta_{2})^{k/2}}\sim\left(\frac{d(\Delta E)}{d(\Delta E_{U}^{(k)})}\right)^{\frac{k}{2}-1}\rightarrow0\ , 
	\end{eqnarray}
	for small $\Delta E$. Thus, we immediately conclude that for any small but fixed $\epsilon$, Gaussian RMT scale $\Delta E_{GUE}^{(k)}$ will be (much) smaller, but not parametrically (in terms of system-size dependence) smaller than $\Delta E_{U}^{(k)}$.

	\begin{figure}[tb]
		\includegraphics[width=1.0\linewidth]{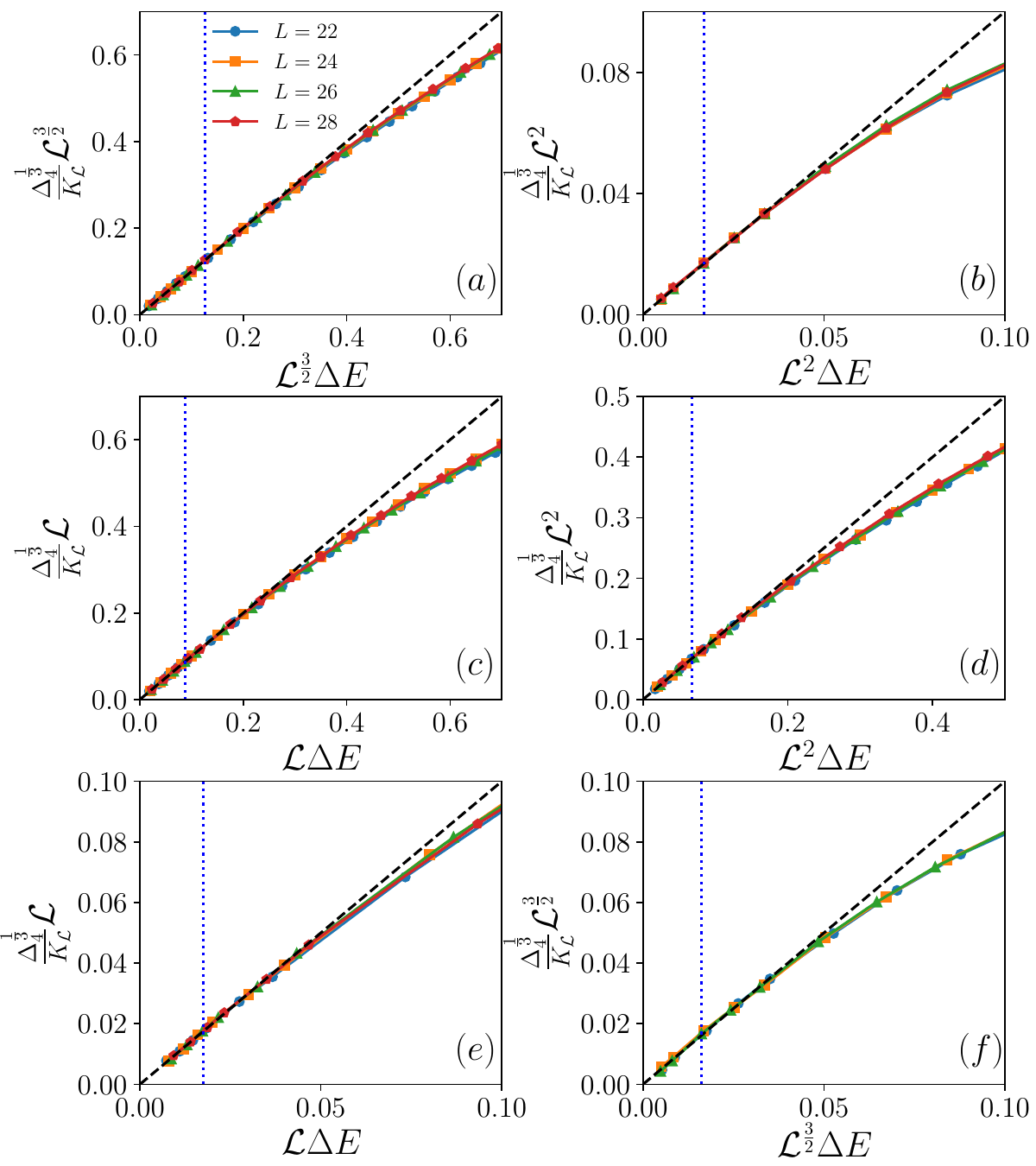}
		\caption{$\frac{\Delta_{4}^{\frac{1}{3}}}{K_{\cal L}}{\cal L}^{p}$ versus ${\cal L}^p \Delta E$ with the best fit values of $K_{\cal L}$ and $p$ to achieve $L$-independence (best data collapse)  for 
			(a) ${\cal A}_{q = \frac{L}{2}}$; (b) ${\cal A}_{q=1}$; (c) ${\cal B}_{q=\frac{L}{2}}$;  (d) ${\cal B}_{q=1}$; (e) ${\cal B}_{q=0}$ and  
			(f) local energy operator ${\cal A}$ \eqref{EQLocalA}. Here ${\cal L} = L/24$ and we use $p=n/2$ with integer $n$ as a tuning parameter.  As a  guide to the eye, inclined dashed black lines  and vertical dotted blue lines   indicate theoretical behavior $\Delta_{k}^{\frac{1}{k-1}}\propto\Delta E$ and  the approximate location of $\Delta E_U$, respectively.}\label{Fig-Delta4-L}
	\end{figure}
	%
	
	This conclusion, combined with the result of \cite{PhysRevLett.128.190601} suggests a non-trivial scaling hierarchy. 
	Indeed, by considering thermalization of the density-wave operators (of the type introduced in Eq. (11) in the main text) in 1D systems, and assuming they are coupled to conserved quantity with the sub-ballistic (e.g.~diffusive) transport, Ref.~\cite{PhysRevLett.128.190601} imposed a bound on the $\Delta E_{GUE}$ scale, 
	\begin{eqnarray}
	\label{bound}
	\Delta E_{GUE}\lesssim \Delta E_T/L\ .  
	\end{eqnarray}
	This scale has a simple physical interpretation of inverse timescale when the expectation value of $\mathcal O$ in some initial out-of-equilibrium state $\ket{\Psi}$ will become of order of quantum fluctuations,
	\begin{eqnarray}
	\langle \Psi|{\mathcal O}(t)|\Psi\rangle \approx e^{-\Delta E_T t} \sim e^{-S/2}\ .
	\end{eqnarray}
	The bound \eqref{bound} is based on the property of Gaussian RMT that the largest eigenvalue of the microcanonically truncated operator will grow with $\Delta E$ as $\sqrt{\Delta E}$. (This is the same as the size of Wigner's semicircle distribution being proportional to square root of the matrix size $\sqrt{d}$.) This behavior is guaranteed, provided 
	$\Delta_2^k$ gives leading contribution to ${\cal M}_{2k}$ \eqref{M2k}, or
	$\Delta_{2k}/(\Delta_2)^{k}\ll 1$ for $k\gg 1$. In other words, for $k\gg 1$, $\Delta E_{GUE}^{(k)}$ obeys \eqref{bound}.
	Combining all together we find $\Delta E_{GUE}^{(k)}\ll \Delta E_{U}^{(k)}$ while both scale with $L$ as
	\begin{eqnarray}
	\label{scaling}
	\Delta E_{T}/L\sim{q^2\over L^3}\ ,
	\end{eqnarray}
	and thus {\it parametrically} smaller than $\Delta E_T$. 
	
	We emphasize, the system size scaling \eqref{scaling} should apply to $\Delta E_{GUE}^{(k)}, \Delta E_{U}^{(k)}$ with  $k\gg 1$, while in our numerical analysis we used $k=4$. We leave open the question of how 
	$\Delta E_U^{(k)}/\Delta E_U^{(l)}$ would scale with the system-size when $k\gg l$. Elucidating this question would be necessary to confirm the hierarchical behavior $\Delta E_{GUE} \ll \Delta E_{U} \ll \Delta E_T$ introduced in the main text.
	

	\subsection*{Additional numerical results}\label{App::Addition}
	{\it System size dependence of $\Delta E_U$.}
	\begin{figure}[tb]
		\includegraphics[width=1.0\linewidth]{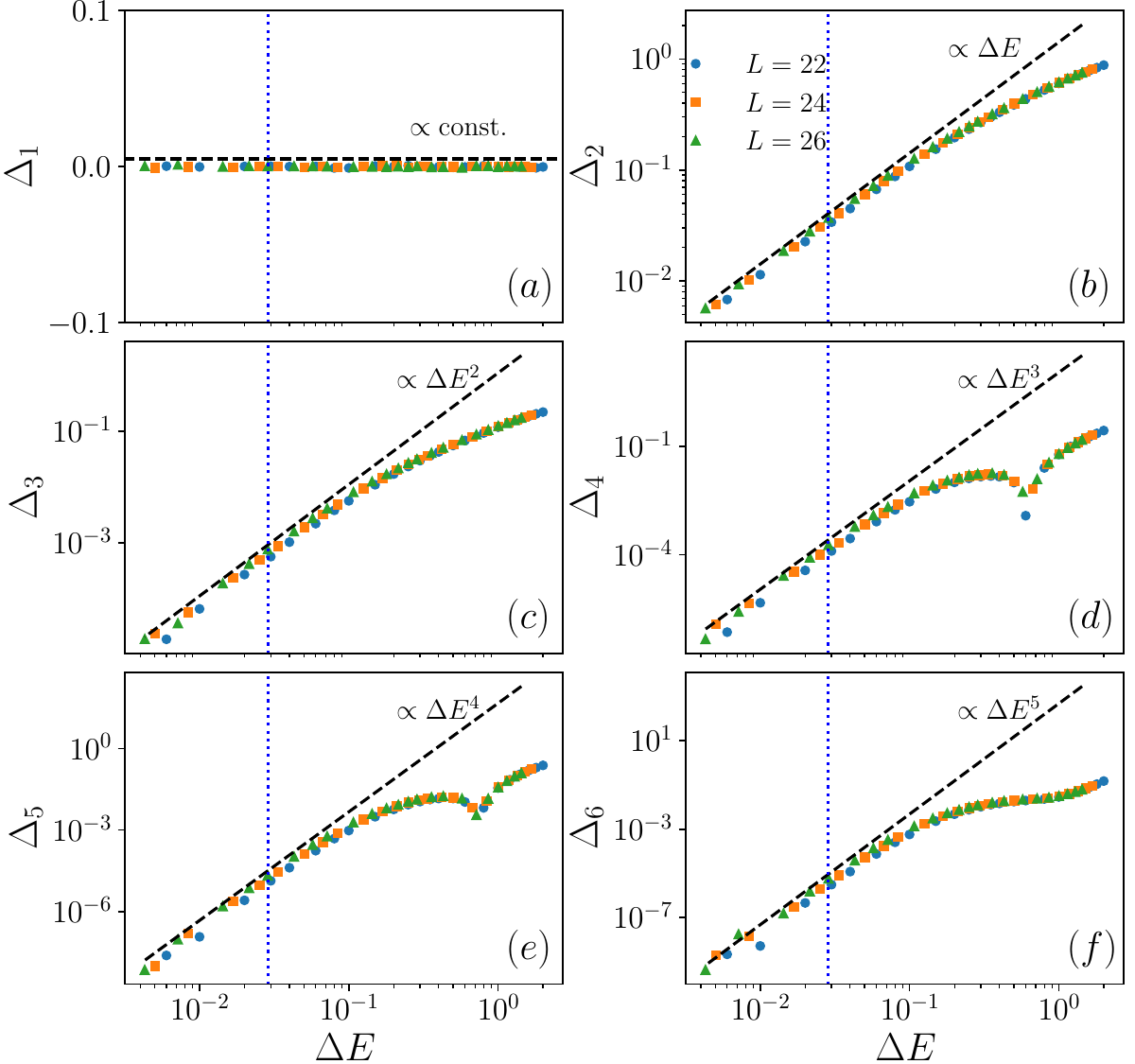}
		\caption{Free cumulants $\Delta_{k}$ versus $\Delta E$ for local operator ${\cal A}$.
			{As a guide to the eye, inclined dashed lines (black) and vertical dotted lines (blue)  indicate  theoretically predicted slope $\ln\Delta_{k}=(k-1)\ln \Delta E+{\rm const}$, and  an approximate location of $\Delta E_U$, respectively.}
		}\label{Fig-LocalH}
	\end{figure}
	Numerically, the procedure to determine $\Delta E_U$ from Eq. \eqref{eq:EUdef} is not very accurate. 
	To study $L$ dependence of $\Delta E_U$ we employ a different approach.
	Considering an operator of interest, we try to find two tuning parameters $K_L$ and $p$ such that the curves of ${\Delta_4^{1/3}L^p\over K_L}$ as a function of $L^p \Delta E$ for different $L$ would approximately be $L$-independent. We do this in two steps. We first consider  $\Delta_{4}^{\frac{1}{3}}$, which grows linearly for small  $\Delta E$, and find $K_L^{-1}$ such that for very small $\Delta E$, $K_L^{-1} \Delta_{4}^{\frac{1}{3}} \approx \Delta E$. Next we find a tuning parameter $p$ to achieve best possible data collapse ($L$-independence) of  $\frac{\Delta_{4}^{\frac{1}{3}}}{K_L}{L}^{p}$
	as a function of ${L}^{p}\Delta E$. We choose $p$ in the form $p=\frac{n}{2}$ with integer $n$. 
	For convenience instead of $L$ we use ${\cal L} = L/24$ and plot $\frac{\Delta_{4}^{\frac{1}{3}}}{K_{\cal L}}{\cal L}^{p}$
	as a function of ${\cal L}^{p}\Delta E$.
	The results are shown in in Fig.\ \ref{Fig-Delta4-L}, where we find $p$ to be positive for all operators considered. This suggests $\Delta E_U \sim L^{-p}$ scaling with positive $p$.
	
	Our results suggest that 
	$\Delta E_{U}\propto {\cal L}^{-p}$ with $p \geq 0$, but the precise value of $p$ is dependent on the observable of interest.
	While a perfect data collapse is difficult to achieve for most observables, the precise finite-size scaling of $\Delta E_U$ remains  unclear. This should not come as a surprise, as the system sizes $L=22,\dots,28$ are likely still too small to observe the asymptotic scaling regime. 
	
	{\it Local operators.}
	Next we consider a local energy density operator ,
	\begin{equation}\label{EQLocalA}
	{\cal A}=\frac{h}{2}(\sigma_{x}^{1}+\sigma_{x}^{2})+\frac{g}{2}(\sigma_{z}^{1}+\sigma_{z}^{2})+J\sigma_{z}^{1}\sigma_{z}^{2}\ .
	\end{equation}
	Numerical results are show  in Fig.~\ref{Fig-LocalH}.
	Similar to the density wave operator, apart from the deviations at extremely small $\Delta E$, $\Delta_k \propto \Delta E^{k-1}$ behavior is  observed below certain energy scale, indicating emergence of unitary symmetry at $\Delta E \le \Delta E_U$. 
	
\begin{figure}[tb]
		\centering
		\includegraphics[width = 1.0\linewidth]{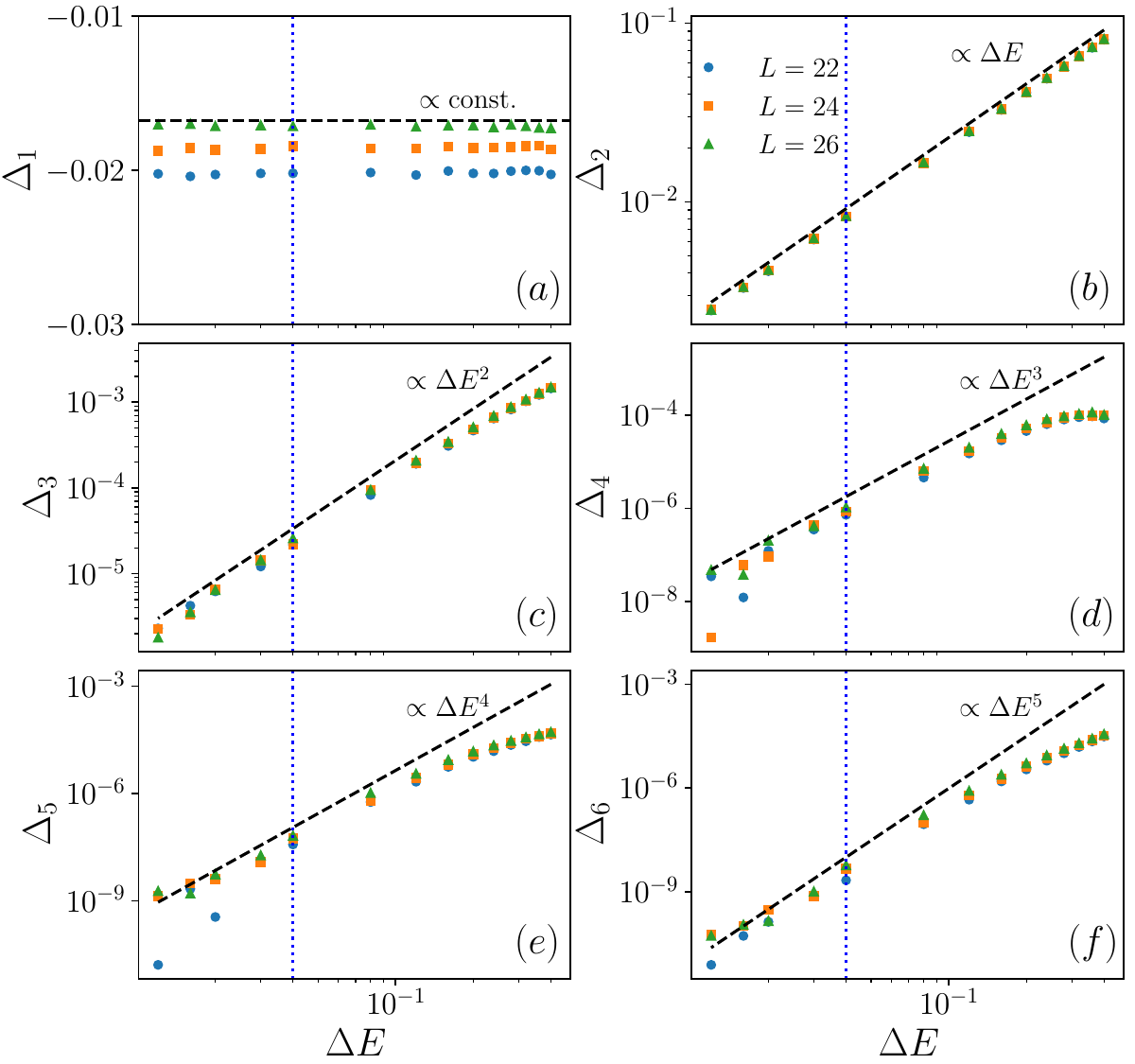}s
		\caption{Similar to Fig.~\ref{Fig-LocalH}, the results for local operator ${\cal B}$. }\label{Fig-B-Local}
	\end{figure}

	Finally in Fig.~\ref{Fig-B-Local}, we study another local operator
	\begin{equation}
	{\cal B}=h\sigma_{z}^{1}-g\sigma_{x}^{1}.
	\end{equation}
	Similarly, $\Delta_k \propto \Delta E^{k-1}$ behavior emerges within a sufficiently small energy window.
	%

	
	{\it Indicator of distance to GUE.\ } 
	{For an operator exhibiting unitary symmetry, 
		Eq.~\eqref{eq-Deltak-DE} implies $\Delta_{k}/(\Delta_{2})^{k/2}=\lambda_k/\lambda^{k/2}_2 \Delta E^{k/2-1}\rightarrow0$ when $\Delta E \to 0$, which indicates emergence of GUE behavior at sufficiently small $\Delta E$. Therefore, the factor $\lambda_k/\lambda^{k/2}_2$ can be regarded as an indicator of distance to GUE at given $\Delta E$.
		In Table. \ref{table1}
		we report the (approximate) values of  $\mu_k=(\lambda^{k/2}_2/\lambda_k)^{1/(k/2-1)}$ for several operators and  different system sizes. Values of $\mu_k$, which have dimension of energy, would indicate the scale where deviation of $k$-th cumulant from the GUE is still very significant. The observed values of $\mu_4$ are of order $0.1-1$, highlighting that GUE is only applicable at scales $\Delta E\ll 1 $.}
	
	
	
	\begin{table}[h!]
		
		\begin{tabular}{||c | c||c | c ||} 
			\hline
			& $\mu_4\ \ $ & & $\mu_4\ \ $ \\ [0.5ex] 
			\hline\hline
			${\cal A}_{L/2}(L=24)$ & 1.0 & ${\cal B}_{L/2}(L=24)$ & 0.8  \\ 
			${\cal A}_{L/2}(L=28)$ & 0.9 &  ${\cal B}_{L/2}(L=28)$ & 0.7  \\
			${\cal A}_{1}(L=24)$ & 0.1 &  ${\cal B}_{1}(L=24)$ & 0.6     \\
			${\cal A}_{1}(L=28)$ & 0.08 &  ${\cal B}_{1}(L=28)$ & 0.5    \\
			[1ex] 
			\hline
		\end{tabular}
		\caption{List of approximated value of $\mu_4$  for different operators and different system sizes.}\label{table1}
	\end{table}

\begin{figure}[tb]
		\centering
		\includegraphics[width=1.0\linewidth]{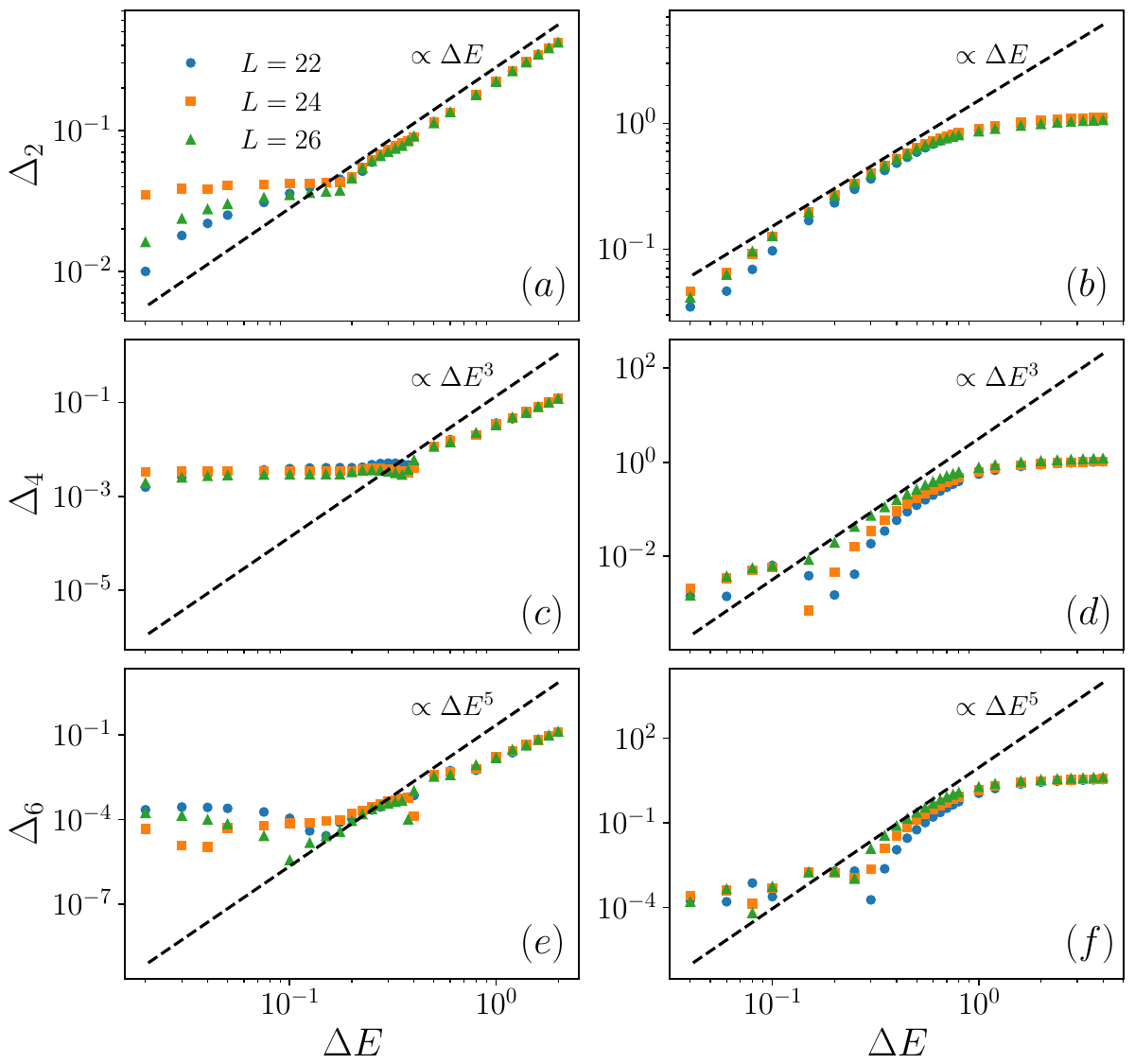}
		\caption{Even cumulants $\Delta_k$ for $k=2,4,6$ as a function of $\Delta E$ in the (integrable) transverse field Ising model \eqref{eq-HIsing-Int}, for the density-wave operator ${\cal A}_q$ with the wave-number $q=\frac{L}{2}$ (panels  [(a),(c),(e)])  and with  $q = 1$ (panels [(b),(d),(f)]). Data is shown for different system sizes $L=22,24,26$. As a guide to the eye, inclined dashed black lines  indicate the slope $\ln\Delta_k=(k-1)\ln \Delta E+{\rm const}$.
		}\label{Fig-DW-Int}
	\end{figure}

	{\it Integrable system.}
	{Complementary to chaotic systems,
		we also consider an integrable model, an Ising model with inhomogeneous transverse field
		\begin{equation}\label{eq-HIsing-Int}
		H=\sum_{\ell=1}^{L}\left(g\sigma_{x}^{\ell}+J\sigma_{z}^{\ell}\sigma_{z}^{\ell+1}\right)+h_{1}\sigma_{x}^{\lfloor\frac{L}{3}\rfloor}+h_{2}\sigma{}_{x}^{\lfloor\frac{2L}{3}\rfloor}\ ,
		\end{equation}
		where $J = 1, g = \frac{\sqrt{5}}{2}, h_1\simeq 0.062, h_2\simeq -0.09$.
		The two defect terms are added to break 
		the translational and reflection symmetries of $H$.
		The observables considered are energy density wave operators
		\begin{equation}
		{\cal A}_{q}=\frac{1}{\sqrt{L}}\sum_{\ell=1}^{L}\cos(\frac{2\pi}{L}q\ell){\cal A}^{\ell},
		\end{equation}
		where
		\begin{equation}
		{\cal A}^{\ell}=\frac{h}{2}(\sigma_{x}^{\ell}+\sigma_{x}^{\ell+1})+J\sigma_{z}^{\ell}\sigma_{z}^{\ell}.
		\end{equation}
		In contrast to chaotic cases, deviation from the $\Delta_k \propto \Delta E^{k-1}$ behavior is clearly visible in Fig. \ref{Fig-DW-Int}, indicating that unitary symmetry does not emerge even at very small scales.}

	\subsection*{Details of the numerical method}
	The key idea of our numerical approach is the expansion of projector operator $P_{\Delta E} = \sum_{|E_m-E_0| < \Delta E/2} \ket{m}\bra{m}$ in terms of Chebyshev polynomials of the Hamiltonian given, see Eq.\ \eqref{Eq::FilterApply} in the main text. 
	In this section, we are going to explain the numerical method in more detail.
	
	In the eigenbasis, $P_{\Delta E}$ can be written as, 
	\begin{align}
	P_{\Delta E}&=\sum_{m}P_{\Delta E}(E_{m})|m\rangle\langle m|\ ,  \label{eq-PDE} \\
	P_{\Delta E}(E)&=\text{rect}\left(\frac{E-E_{0}}{\Delta E}\right)\ ,  
	\end{align}
	where $\text{rect}(x)$ indicates the rectangular function defined as,
	\begin{equation}
	\text{rect}(x)=\begin{cases}
	0 & \text{if }|x|\ge\frac{1}{2}\\
	1 & \text{if }|x|<\frac{1}{2}
	\end{cases}\ .
	\end{equation}
	The rectangular function can be expanded in terms of Chebyshev polynomials of the first kind (denoted by $T_k(x)$),  
	\begin{equation}\label{eq-PE}
	P_{\Delta E}(E) = \sum_{k=0}^{\infty} C_k T_k \left(\frac{E-b}{a}\right),
	\end{equation}
	where  $a = 
	(E_\text{max}-E_\text{min})/2$, $b = (E_\text{max} + E_\text{min})/2$. The coefficients $C_k$ are written as 
	\begin{equation}\label{eq-C0}
	C_{0}=\frac{1}{\pi}\left[\arcsin\left(\frac{\frac{\Delta E}{2}+b-E_{0}}{a}\right)-\arcsin\left(\frac{-\frac{\Delta E}{2}+b-E_{0}}{a}\right)\right]\ ,
	\end{equation}
	and 
	\begin{align}\label{eq-Ck}
	C_k(k\ge 1)  & = \frac{2(-1)^{k+1}}{\pi k}\biggl[ 
	\sin\biggl(k\arccos\biggl(\frac{\frac{\Delta E}{2}+b-E_0}{a}\biggl)\biggl) \nonumber 
	\\
	& -\sin\biggl(k\arccos\biggl(\frac{-\frac{\Delta E}{2}+b-E_0}{a}\biggl)\biggl)\biggl] 
	\ .
	\end{align}

	In practice, the infinite series in Eq.\ \eqref{eq-PE} needs to be truncated to finite order $N_\text{tr}$,
	\begin{equation}\label{eq-PEM}
	P_{\Delta E}(E) \simeq \sum_{k=0}^{N_\text{tr}} C_k T_k \left(\frac{E-b}{a}\right)\ .
	\end{equation}
	In the following, we denote $N_\text{tr}$ by $N$ for simplicity.
	$P_{\Delta E}$ finally can be approximated as
	\begin{equation}\label{Eq::FilterApply}
	P_{\Delta E}\simeq P_{\Delta E}^{(N)}\equiv\sum_{k=0}^{N}C_{k}T_{k}\left(\frac{H-b}{a}\right)\ .
	\end{equation}
	Applying the above equation to a Gaussian random state 
	\begin{equation}
	|\psi\rangle = \sum_{i=1} ^ {D} c_i |i\rangle\ , 
	\end{equation}
	where $c_i$ is a Gaussian random number and $\ket{i}$ denote the computational basis, the 
	$k$-th moment of an operator $\cal O$ can be approximated by relying on quantum typicality, 
	\begin{equation}\label{eq-Mk0}
	{\cal M}_{k}\simeq\frac{\langle\psi|(P_{\Delta E}^{(N)}{\cal O}P_{\Delta E}^{(N)})^{k}|\psi\rangle}{\langle\psi|P_{\Delta E}^{(N)}|\psi\rangle}\ . 
	\end{equation}
	Moreover, making use of $\left(P_{\Delta E}^{(N)}\right)^{2}\simeq P_{\Delta E}^{(N)}$,  Eq.\ \eqref{eq-Mk0} is simplified as 
	\begin{equation}\label{eq-Mk}
	{\cal M}_{k}\simeq{\cal M}_{k}^{\text{typ}}\equiv\frac{\langle\psi|P_{\Delta E}^{(N)}({\cal O}P_{\Delta E}^{(N)})^{k}|\psi\rangle}{\langle\psi|P_{\Delta E}^{(N)}|\psi\rangle}\ , 
	\end{equation}
	where one only needs to apply the energy filter $P_{\Delta E}^{(N)}$ for $k+1$ times, instead of $2k$ times in the original expression \eqref{eq-Mk0}.
	In the numerical simulations, to further reduce statistical errors, an additional average is taken over $N_\text{typ}$ different realization of random state, 
	\begin{equation}
	\overline{{\cal M}_{k}^{\text{typ}}}=\frac{1}{N_{\text{typ}}}\sum_{n=1}^{N_{\text{typ}}}\frac{\langle\psi_{n}|P_{\Delta E}^{(N)}({\cal O}P_{\Delta E}^{(N)})^{k}|\psi_{n}\rangle}{\langle\psi_{n}|P_{\Delta E}^{(N)}|\psi_{n}\rangle}\ .
	\end{equation}
	where $|\psi_n\rangle$ indicates an individual realization of a Gaussian random state.
	The Chebyshev expansion becomes more accurate if one increases the number of terms in the expansion $N$, but the simulation time is proportional to $N$, hence
	larger $N$ will result in a longer simulation time.  In our numerical simulations $N=100a\frac{2\pi}{\Delta E}$ which is found to yield high accuracy of results.
	\subsection*{Error analysis}
	\begin{figure}[tb]
		\includegraphics[width=1.0\linewidth]{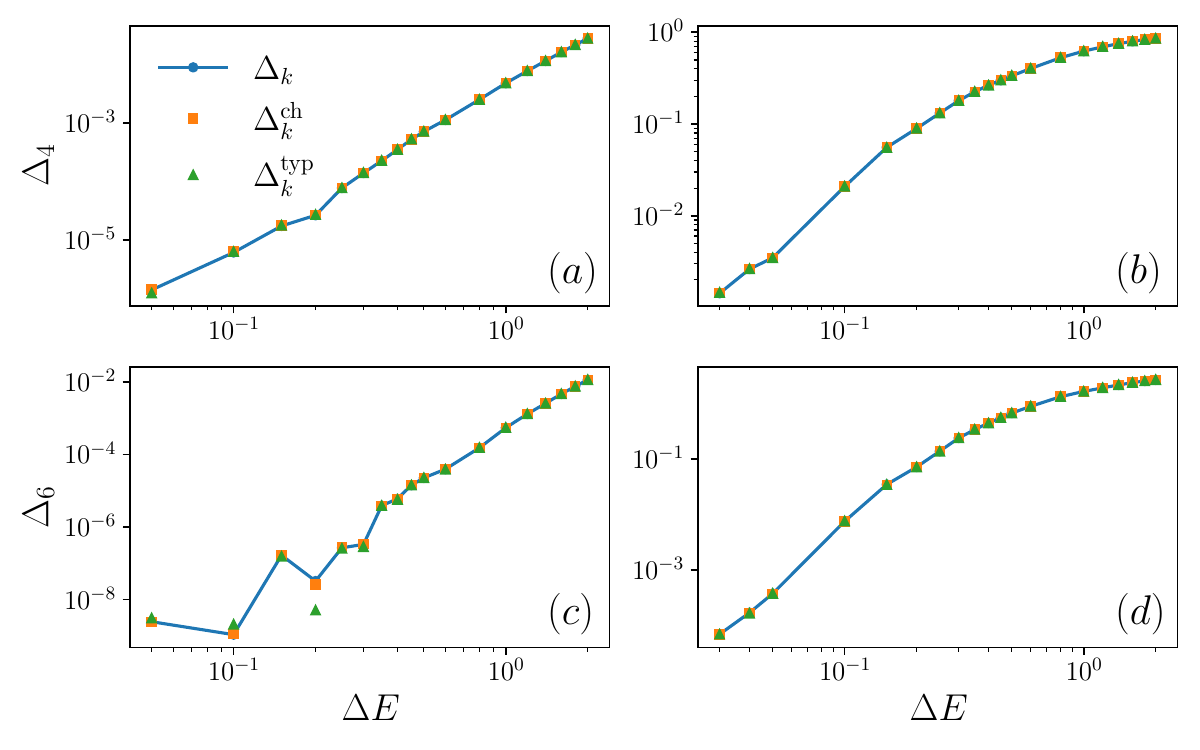}
		\caption{Comparison of $\Delta^\text{typ}_k$ (green triangle, averaged over $2^{12}$ random states),  $\Delta^\text{ch}_k$ (orange square) and $\Delta_k$ (blue circle with solid line, calculated by ED) for operator: [(a) (c)] ${\cal A}_{q=\frac{L}{2}}$ and [(b) (d)] ${\cal A}_{q=1}$ for system size $L=16$.}\label{Fig-Check1}
	\end{figure}

	In this section, we perform an error analysis of our typicality-based numerical method. 
	The total error in our approach comes from two different sources: (i) a truncation error $\varepsilon_\text{tr}$ [due to the finite order $N$ in the Chebyshev expansion \eqref{Eq::FilterApply}], and (ii) a typicality error $\varepsilon_\text{typ}$ [due to approximating the trace by a random state in Eq. \eqref{eq-Mk0}].
	The typicality error $\varepsilon_\text{typ}$ is easy to estimate. It  scales as $\varepsilon_\text{typ} \sim (N_\text{typ}d)^{-1/2}$, where $d$ is the number of states in the energy window. In the numerical simulations, $N_\text{typ}$ is chosen according to the system size, $N_\text{typ} \propto 2^{-L}$, which then assures similar accuracy for different $L$.
	More precisely, the typicality error for the approximation of ${\cal M}_k$ is given by 
	\begin{equation}
	\varepsilon_{\text{typ}}({\cal M}_{k})=\sqrt{\frac{{\cal M}_{2k}-{\cal M}_{k}^{2}}{N_{\text{typ}}d}}\ .
	\end{equation}
	The relative (typicality) error is given by
	\begin{equation}
	\frac{\varepsilon_{\text{typ}}({\cal M}_{k})}{|{\cal M}_{k}|}=\sqrt{\frac{{\cal M}_{2k}-{\cal M}_{k}^{2}}{{\cal M}_{k}^{2}}}\frac{1}{\sqrt{N_{\text{typ}}d}}\ .
	\end{equation}
	We will readily see that it depends on the  order $k$. 
	To estimate the relative error of even moments, let's consider the case where all odd moments vanish.  It is known that if an operator can be described by a GOE random matrix, one has
	\begin{equation}\label{eq-Mk-GOE}
	{\cal M}_{2k}=\Lambda_{k}({\cal M}_{2})^{k}\ , 
	\end{equation}
	where $\Lambda_k = \frac{(2k)!}{(k+1)!k!}$ is the Catalan number.
	Within a small energy window $\Delta E \ll \Delta E_U$,  {$k$-th cumulant decays as $\Delta_{k}\propto{(\Delta E/\Delta E_{U})}^{k-1}$. In this case, Eq.~\eqref{eq-Mk-GOE} holds approximately, at least with respect to scaling}. Therefore, one has
	\begin{equation}
	\frac{{\cal M}_{2k}-{\cal M}_{k}^{2}}{{\cal M}_{k}^{2}}\sim\frac{\Lambda_{k}}{\Lambda_{\frac{k}{2}}^{2}}\sim k^{\frac{3}{2}}\ ,\quad \text{for even } k \gg 2\ .
	\end{equation}
	As a result one has the following estimation for even moments
	\begin{equation}
	\frac{\varepsilon_{\text{typ}}({\cal M}_{k})}{|{\cal M}_{k}|}\sim k^{\frac{3}{4}}\frac{1}{\sqrt{ N_{\text{typ}}d}}\ ,
	\end{equation}
	indicating that the relative error of ${\cal M}_k$ increases as a power law in $k$. For the density-wave operators with non-zero wave number, odd moments approximately vanish. So in our numerical simulation, we neglect all the odd moments and only even moments are considered.
	For all other operators, for which the odd moments are not negligibly small,  the ratio $\frac{{\cal M}_{2k}-{\cal M}_{k}^{2}}{{\cal M}_{k}^{2}}$ is operator-dependent.  Usually we expect that $\frac{{\cal M}_{2k}-{\cal M}_{k}^{2}}{{\cal M}_{k}^{2}}$ increase with $k$, which leads to a larger relative error for higher moments.
	It should be mentioned here that, as free cumulants $\Delta_k$ are calculated by making use of moments ${\cal M}_k$, their relative error are in general larger than that of ${\cal M}_k$.
	
	\begin{figure}[tb]
		\includegraphics[width=1.0\linewidth]{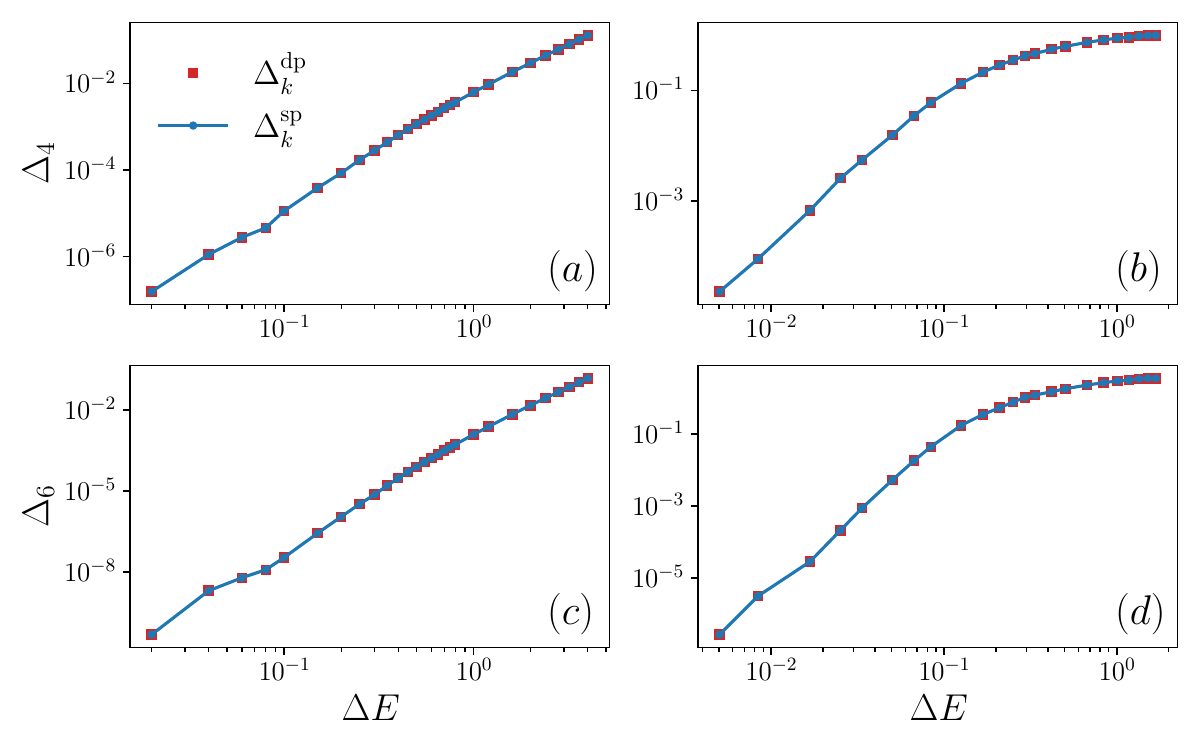}
		\caption{Comparison of result of $\Delta_k$ from single precision $\Delta^\text{sp}_k$ (blue circle)  and double precision $\Delta^\text{dp}_k$ (red square) simulations, for operator: [(a) (c)] ${\cal A}_{q=\frac{L}{2}}$ and [(b) (d)] ${\cal A}_{q=1}$ for system size $L=24$. The result is obtained from a same random state.}\label{Fig-Check32}
	\end{figure}

	The truncation error $\varepsilon_\text{tr}$ depends on the number of Chebyshev polynomials we keep in the expansion of Eq. \eqref{Eq::FilterApply}. Different from the typicality error,  an analytical expression of truncation error is very hard to derive.
	We try to estimate $\varepsilon_\text{tr}$ numerically instead.
	To this end, we consider 
	\begin{equation}
	{\cal M}_{k}^{\text{ch}}\equiv\frac{\text{Tr}\left[P_{\Delta E}^{(N)}({\cal O}P_{\Delta E}^{(N)})^{k}\right]}{\text{Tr}\left[P_{\Delta E}^{(N)}\right]}\ .
	\end{equation}
	It can be easily seen that
	${\cal M}_{k}^{\text{ch}} \to {\cal M}_k$ for $N\rightarrow \infty$ and
	$\overline{{\cal M}_{k}^{\text{typ}}} \to {\cal M}_{k}^{\text{ch}}$
	if the Hilbert-space dimension is sufficiently large or if the results are averaged over many realizations of random state. Here $\overline{{\cal M}_{k}^{\text{typ}}}$ indicates that the trace is approximated by random states.
	We denote the cumulants calculated by $\overline{{\cal M}_{k}^{\text{typ}}}$ and ${\cal M}_{k}^{\text{ch}}$ as $\Delta^\text{typ}_k$ and $\Delta^\text{ch}_k$, respectively. The error $\varepsilon_\text{typ}$ and $\varepsilon_\text{tr}$ can be probed by
	\begin{equation}
	\varepsilon_{\text{tr}}\propto|\Delta_{k}^{\text{ch}}-\Delta_{k}|\ ,\quad\varepsilon_{\text{typ}}\propto|\Delta_{k}^{\text{ch}}-\Delta_{k}^{\text{typ}}|\ .
	\end{equation}
	
	We compare $\Delta^\text{typ}_k$, $\Delta^\text{ch}_k$ and  $\Delta_k$ for system size $L=16$ in Fig.\ \ref{Fig-Check1} .
	For the operators considered, a neat agreement between $\Delta_{k}^{\text{ch}}$ and $\Delta_{k}$ can be seen, indicating a small truncation error $\varepsilon_\text{tr}$. 
	$\Delta_{k}^{\text{typ}}$ (calculated by averaging over $2^{12}$ random states) also remains very close to $\Delta_{k}^{\text{ch}}$ for almost all $
	\Delta E$, indicating a small typicality error $\varepsilon_\text{typ}$. Deviations can be observed for $\Delta E \rightarrow 0$, especially for $\Delta_6$ of ${\cal A}_{q=\frac{L}{2}}$ [Fig. \ref{Fig-Check1}~(c)] when  the number of states within the  energy window  plunges below  $d\lesssim 1000$. 
	
	It should be mentioned here that all our numerical simulations with the typicality-based method are done in single precision. The accuracy of single precision simulation is checked in several observables (see Fig. \ref{Fig-Check32}), where we find that the difference between results from single and double precision simulations are neglectably small.

\end{document}